\documentclass[10pt, conference, letterpaper]{IEEEtran}
\IEEEoverridecommandlockouts
\usepackage{cite}
\usepackage{amsfonts}
\usepackage{amsthm}
\usepackage[utf8]{inputenc}
\usepackage{tikz,forest}
\usepackage{comment}
\usepackage{algorithm}
\usepackage[noend]{algpseudocode}
\usepackage[english]{babel}
\usepackage{url}
\usetikzlibrary{shapes.geometric, arrows}
\usepackage{paralist}
\usepackage{caption}
\usepackage{subcaption}
\usepackage{xspace}
\usepackage{footnote}
\usepackage{epigraph}
\makesavenoteenv{algorithm}

\algrenewcommand\algorithmicindent{1.0em}%

\newtheorem{theorem}{Theorem}

\newtheorem{lemma}[theorem]{Lemma}

\theoremstyle{remark}

\theoremstyle{definition}
\newtheorem{definition}{Definition}

\usepackage{mathtools}

\allowdisplaybreaks

\newcommand{\abs}[1]{\left\vert#1\right\vert}
\newcommand{\set}[1]{\left\{#1\right\}}
\newcommand{\reals}{\mathbb{R}}

\newcommand{\naturals}{\mathbb{N}}

\DeclareMathOperator*{\argmin}{arg\,min}

\newcommand{\topalg}{Top}
\newcommand{\maxalg}{Max}
\newcommand{\levelalg}{Level}
\newcommand{\numswitches}{n}
\newcommand{\switchset}{\mathcal{S}}
\newcommand{\switch}{s}
\newcommand{\serverset}{\mathcal{W}}
\newcommand{\server}{w}
\newcommand{\network}{T}
\newcommand{\vertexset}{V}
\newcommand{\linkset}{E}
\newcommand{\rootswitch}{r}
\newcommand{\link}{e}
\newcommand{\load}{L}
\newcommand{\parent}{p}
\newcommand{\numblue}{\ensuremath{k}}
\newcommand{\blueset}{U}

\newcommand{\blue}{B}
\newcommand{\red}{R}
\newcommand{\weight}{\omega}
\newcommand{\rate}{\tau}
\newcommand{\aggcap}{a}
\newcommand{\msg}{\mathrm{msg}}
\newcommand{\util}{\ensuremath{\psi}}
\newcommand{\childnum}{C}
\newcommand{\combic}{C-BIC}
\newcommand{\msgsize}{M}
\newcommand{\msgcost}{\psi}
\newcommand{\destination}{d}
\newcommand{\alg}{SMC}

\newcommand{\reduce}{Reduce\xspace}

\DeclareMathOperator{\minsplit}{mSplit}
\DeclareMathOperator{\mincost}{mCost}
\DeclareMathOperator{\nodecolor}{color}

\newcommand{\wcapp}{WC}
\newcommand{\psapp}{PS}

\newcommand{\mbn}{X}
\newcommand{\up}{\beta}

\algnewcommand{\algorithmicand}{\textbf{ and }}
\algnewcommand{\algorithmicor}{\textbf{ or }}
\algnewcommand{\OR}{\algorithmicor}
\algnewcommand{\AND}{\algorithmicand}

\newcommand{\Availability}{\Lambda}

\newcommand{\Gather}{Gather}
\newcommand{\Color}{Color}
\newcommand{\alggather}{\alg-\Gather}
\newcommand{\algcolor}{\alg-\Color}

\tikzset{
  treenode/.style = {align=center, inner sep=0pt, text centered},
  node_b/.style = {treenode, circle, white, draw=black, font=\Huge, fill=blue, text width=3em},
  node_r/.style = {treenode, circle, red, draw=red, font=\Huge, text width=3em, very thick},
  node_ws/.style = {treenode, circle, draw=white, font=\Huge, text width=3em, very thick},
  node_p/.style = {treenode, rectangle, draw=black, font=\Huge, minimum height = 1.1cm, text width=3em, very thick},
  node_h/.style = {treenode, rectangle, white, draw=black, font=\Huge, minimum height = 1.1cm, text width=3em, fill=gray, very thick},
  node_wh/.style = {treenode, rectangle, draw=white, font=\Huge, minimum height = 1.1cm, text width=3em, very thick},
  edge from parent/.style = {font=\Huge, line width=1mm, draw, <-, >=stealth'},
  level 2/.style = {sibling distance=40mm},
  level 3/.style = {sibling distance=20mm},
  level/.style = {level distance=2.5cm},
}

\usepackage{soul}

\newcommand{\revision}[1]{{#1}}

\begin{document}

\date{}

\title{Constrained In-network Computing \\ with Low Congestion in Datacenter Networks
}

\author{
{\rm Raz Segal, Chen Avin, and Gabriel Scalosub}\\
School of Electrical and Computer Engineering\\
Ben-Gurion University of the Negev, Israel\\
razseg@post.bgu.ac.il, avin@bgu.ac.il, sgabriel@bgu.ac.il
} 

\maketitle
\begin{abstract}


Distributed computing has become a common practice nowadays,
where recent focus has been given to the usage of smart networking devices with in-network computing capabilities.
State-of-the-art switches with near-line rate computing and aggregation capabilities enable acceleration and improved performance for various modern applications like big data analytics and large-scale distributed and federated machine learning.

In this paper, we formulate and study the theoretical algorithmic foundations of such approaches, and focus on how to deploy and use constrained in-network computing capabilities within the data center.
We focus our attention on reducing the network congestion, i.e., the most congested link in the network, while supporting the given workload(s).
We present an efficient optimal algorithm for tree-like network topologies and show that our solution provides as much as an x13 improvement over common alternative approaches.
In particular, our results show that having merely a small fraction of network devices that support in-network aggregation can significantly reduce the network congestion, both for single and multiple workloads.
\end{abstract}




\section{Introduction}
\label{sec:introduction}
As online applications and services increase in popularity, distributed data processing capabilities and datacenter networks
have become a major part of the infrastructure of modern society. 
Moreover, due to the vast growth in the amount of data processed by such applications, recent work shows that the {\em bottleneck} for efficient distributed computation is now the underlying communication {\em network} and not the computational capabilities at the servers~\cite{chowdhury11managing,mai14netagg,viswanathan20network}, as was traditionally the case.

For example, distributed machine learning (ML) tasks, which are the driving force behind some of the most exciting technological developments of recent years, are significantly constrained by such bottlenecks~\cite{xu20compressed}.
Frequently, communication-intensive and network-wide operations like {\em AllReduce} are essential for such applications to sustain the ever-increasing volumes of data they have to process. 
Other examples are scenarios giving rise to the {\em incast} problem~\cite{alizadeh10dctcp,wu13ictcp} arising also in Big Data applications, e.g., within MapReduce frameworks.


In an effort to improve the performance of such tasks, a recent line of work, both by academia and industry, proposed the usage of {\em in-network computing}~\cite{ports19when,sapio17innetwork,costa12camdoop,graham20sharp}.
This approach tries to offload as much of the computation as possible 
onto ``smart'' networking devices achieving two goals:
\begin{inparaenum}[(i)]
\item possibly reducing 
the amount of data that traverses the network, and
\item reducing or even eliminating some of the computational tasks from servers and end hosts.
\end{inparaenum}
By that, in-network computing aims to significantly improve performance and cost.

This effort is bearing fruit and cutting-edge networking devices like switches and SmartNICs actually perform local computation on streams of traffic, like reduce operations, even at line rate \cite{graham20sharp,gebara21innetwork}.
By using SDN and programmable network elements (e.g., P4)~\cite{bosshart14p4}, such in-network {\em computing devices} are being deployed, and have been shown to greatly improve both networks, and applications, performance, as well as resource usage efficiency~\cite{graham20sharp,gebara21innetwork}.

As there is (probably) no free lunch~\cite{wolpert1997no}
when using in-network computing,
deploying 
such capable devices 
in a network comes at a cost (e.g., usage of computing resources, power consumption, or availability).
Hence, such capabilities might not be ubiquitous throughout the network, or at all times, or for every workload.
For example, when such a service is bundled in a service-level agreement (SLA), or when multiple tenants and multiple workloads
call for such in-network computation abilities, it might be that the available resources that are required to support such in-network computation might not be sufficient for satisfying all pending requirements.

In this work, we focus our attention on
the task of {\em data aggregation} as it occurs in, e.g., MapReduce frameworks, or distributed machine learning frameworks making use of, e.g., a parameter server, or gradient aggregation and distribution.
We study such in-network computing paradigms in tree-based (overlay) topologies
consisting of a tree network of switches, each connected to some number of servers (e.g., switches can be viewed as Top-of-Rack switches).\footnote{Such tree topologies are common as  a virtual overlay over a physical network or as sub-topologies in a data center.}
Our goal is to perform a {\em Reduce} operation, where the data aggregated from all servers
should reach
a special {\em destination} server $d$ (which can be logically viewed as simply the root switch).
It should be noted that tree-based topologies as the one used in our model lay at the core of various popular architectures for distributed machine-learning use cases, implementing, e.g., AllReduce operations~\cite{nvidia19doubletree,sanders2009two,gebara21innetwork}.

We consider the {\em constrained in-network processing} problem~\cite{segal2021soar}, where
we have at our disposal a limited {\em budget} of $\numblue$ aggregation switches, which we can deploy (or activate) in some $\numblue$ locations throughout the network.
Our objective in this work is to minimize the {\em network congestion}, i.e., minimizing the {\em most congested link} throughout the network, where link congestion if defined as the ratio between the number of messages traversing the link (i.e., the link load) and the rate of the link.
Minimizing congestion is notably a key objective in networking, as it bears significant consequences for network and applications performance alike~\cite{banner07multipath,racke08optimal,gainaru15scheduling,bhatele15identifying,bansal15minimum,%
avin19demand,gao20congestion}.

We assume
each aggregating switch deployed in the network provides the ability of aggregating multiple incoming messages onto a {\em single} outgoing message.
For cases where all switches can perform aggregation, one obtains the minimum congestion possible (as each link carries a single message).
On the other extreme, when none of the switches has aggregation capabilities, congestion is extremely high, since essentially all messages must traverse the very few links entering the root.

However, for non extremal values of $\numblue$, finding the optimal placement of a limited number of aggregation switches so as to minimize network congestion, is not a trivial task, even for trees, which is the case considered in this work.
This is due to the fact that such an optimal placement of aggregation switches is affected by various network and workload factors, including
the specific tree topology,
the rates of the links,
the load distribution at the servers,
and the availability of resources for supporting such aggregation at the switches.
Nevertheless, we present {\em an optimal algorithm} for performing such placement.
Addtionally, our results show that
placing relatively few aggregation nodes may
drastically reduce network congestion, if judiciously placed in the proper locations.


Our model and results seem to be especially tailored for cloud environments, where providers may offer in-network aggregation with congestion guarantees as part of their business offerings.
This can be viewed as part of their Network-as-a-Service (NaaS) suite, allowing the dynamic allocation, and re-allocation, of in-network computing capabilities {\em on-demand}.

\subsection{Our Contribution}

We formulate the {\em Congestion-Minimization with Bounded In-network Computing} (\combic) problem, and present an optimal and time efficient algorithm for solving the 
problem for a single workload on tree networks with heterogeneous link rates.
Such topologies are common in datacenter networks, e.g., fat-tree topologies \cite{al2008scalable}.
Our solution uses a hybrid search-and-dynamic-programming approach.

We further extend our framework to support {\em multiple tenants/workloads}, and adapt our algorithms to settings where workloads arrive in an {\em online} fashion.
In these settings each switch may support a limited number of workloads, according to its {\em aggregation capacity}.
Each new workload may use (some) in-network aggregation capabilities, and the aggregation capacities of the switches should be carefully allocated.

We discuss and present various properties of our resulting solutions, and evaluate their performance for various server load distribution, network sizes, workload arrivals, aggregation capacities, and network characteristics.
In our study, we further consider two main {\em use cases}:
\begin{inparaenum}[(i)]
\item MapReduce (using word-count as an illustration), and
\item gradient aggregation for distributed machine learning.
\end{inparaenum}
We further show the benefits of using our algorithm when compared with several natural allocation strategies.
Our results indicate that a small fraction of aggregation switches can already significantly reduce the network congestion in data aggregation tasks.

The paper is structured as follows. 
In Sec.~\ref{sec:model} we introduce our formal system model.
Sec.~\ref{sec:example} provides a motivating example highlighting various aspects of the \combic\ problem.
Sec.~\ref{sec:algorithm} presents an overview of our optimal algorithm \alg\, and the main theoretical results. 
We evaluate our algorithm experimentally in Sec.~\ref{sec:evaluation}.
We conclude the paper with related work and discussion in Secs. \ref{sec:relatedwork} and \ref{sec:disussion_future_work}, respectively.
We note that due to space constraints, we provide merely proof sketches for some of the proofs.
%


 

\section{Preliminaries \& System Model}\label{sec:model}

We consider a system comprising a set of $\numswitches$ switches $\switchset$, a set of servers (workers) $\serverset$, and a special destination server $\destination \notin \serverset$.
We assume there exists a pre-specified {\em root} switch $\rootswitch \in \switchset$, and a {\em weighted tree network}
$\network=(\vertexset ,\linkset,\weight)$,
where
$\vertexset=\switchset \cup \set{\destination}$ and $\linkset=\linkset' \cup \set{(\rootswitch,\destination)}$ for some $\linkset' \subseteq \switchset^2$ forming a tree over the set of switches $\switchset$. 
Let $\weight:\linkset \mapsto \reals^+$ be the rate function of the links (in message per second).
For $e\in \linkset$ let $\rate(e)=\frac{1}{\weight(e)}$.
The tree $\network$ thus consists of the underlying network topology connecting the switches, and connecting the root $\rootswitch$ to the destination $\destination$.

We further assume that all links in $\linkset$ are directed towards $\destination$.
In particular, every switch $\switch \in \switchset$ has a unique {\em parent} switch $\parent(\switch) \in \switchset$ defined as the neighbor of $\switch$ on the unique path from $\switch$ to the $\destination$.
In such a case we say $\switch$ is a {\em child} of $\parent(\switch)$, and we let $\childnum(\switch)$ denote the number of children of switch $\switch$.  

We assume each server $\server \in \serverset$ is connected to a single switch $\switch(\server) \in \switchset$, and let $\load:\switchset \mapsto \naturals$ be the function matching each switch $\switch$ with the number of servers conected to $\switch$.
We refer to $\load$ as the {\em network load}.
Each server $\server$ produces a single message, which is forwarded to $\switch(\server)$, where we assume every message has size at most $\msgsize$, for some (large enough) constant $\msgsize$.
Each switch $\switch$ can be of one of two types, or operates at one of two modes:
\begin{enumerate}[(i)]
\item an {\em aggregating} switch (blue), which can aggregate messages arriving from its children (each of size at most $\msgsize$), to a {\em single} message (also of size at most $\msgsize$) and forwards it to its parent switch $\parent(\switch)$, or
\item a {\em non-aggregating} switch (red), which cannot aggregate messages, and simply forwards each message arriving from any of its children to its parent switch $\parent(\switch)$.
\end{enumerate}
We denote by $\Availability \subseteq \switchset$ the set of switches that are {\em available}
as aggregation switches.
Our view of aggregating switches is applicable to devices which compute, e.g., separable functions~\cite{mosk2006computing}.
In particular, this holds true for aggregation functions computing, e.g., the average, or sum, of the values contained in the messages being sent by the servers.

In what follows we will be referring to aggregating switches as {\em blue} nodes in $\network$, and to non-aggregating switches as {\em red} nodes in $\network$.
Our {\em budget} is denoted by a non-negative integer $\numblue$, which serves as an upper bound on the number of blue nodes allowed in $\network$.
We will usually refer to $\blueset \subseteq \Availability$ as the set of blue nodes in $\network$ and require that $\abs{\blueset} \le \numblue$. 

Given a weighted tree network $\network=(\vertexset,\linkset,\weight)$ with a network load $\load:\switchset \mapsto \naturals$, and a set of blue nodes $\blueset \subseteq \Availability$, we consider a simple \reduce operation on $T$ as detailed in Algorithm \ref{alg:reduce}.
Every switch in the tree processes all messages received from its children and forwards message(s) to its parent.
Every blue node (i.e., a node in $\blueset$) is an aggregation switch and all other switches (i.e., nodes not in $\blueset$) are non-aggregation switches.
The operation ends when the {\em destination} receives the overall (possibly aggregated) information from all the nodes that have a strictly positive load.

\begin{algorithm}[t]
\caption[Algorithm]{\reduce$(\network,\load,\blueset)$}
\label{alg:reduce}
\begin{algorithmic}[1]
\Require A tree $\network$, A network load $\load$, A set of blue node $\blueset$
\Ensure An aggregate information at destination $\destination$
\State For each node $v$ in \network~do:
\While{not received all messages from all children}
        \State process incoming message (by switch type: $\blue, \red$)  
        \State if needed send message to $\parent(v)$ (by switch type: $\blue, \red$)
\EndWhile
\end{algorithmic}
\end{algorithm}

For every link $\link=(\switch,\parent(\switch))$ in $\linkset$, we then define the {\em link load}, $\msg_{\link}(\network,\load,\blueset)$, as the number of messages traversing link $\link$, given the \reduce operation on $\network$, $\load$, and $\blueset$.
we further define the {\em link congestion} $\util_e(\network,\load,\blueset) = \msg_\link(\network,\load,\blueset) \cdot \rate(\link)$,
and refer to
\begin{align}
\label{eq:msgcost_def}
\msgcost(\network,\load,\blueset)=\max_{\link \in \network}\set{ \util_\link(\network,\load,\blueset)}
\end{align}
as the {\em network congestion}.
Our work considers the Congestion-minimization with Bounded In-network Computing (\combic) problem, which aims at minimizing the network congestion, formally defined as follows.

\begin{definition}[\combic]
Given a weighted tree network $\network=(\vertexset,\linkset,\weight)$, a network load $\load:\switchset \mapsto \naturals$, a set of available switches $\Availability$, and a budget $\numblue$, the {\em Congestion-minimization with Bounded In-network Computing} (\combic) problem is finding a set of switches $\blueset \subseteq \Availability$ of size at most $\numblue$ that minimizes the network congestion $\msgcost(\network,\load,\blueset)$. Formally,
\begin{align}\label{eq:mindef}
    \mbox{\combic}(\network,\load, \Availability, \numblue)
    =\arg\min_{\substack{\blueset \subseteq \Availability \\ \abs{\blueset}=\numblue}} \msgcost(\network,\load,\blueset)
\end{align}
\end{definition}

In trying to solve the \combic\ problem, one may use a brute-force approach, and enumerate over all all possible subsets of $\Availability$ of size $k$.
This may work well for a small constant $k$, but it becomes quickly intractable for arbitrary values of $k$.
In what follows we will describe and discuss our {\em efficient} solution, \alg, to the \combic\ problem.

\section{Motivating Example}
\label{sec:example}
We now turn to consider a motivating example highlighting the fact that simple, yet reasonable, approaches might fall short of finding an optimal solution to the \combic\ problem.
Specifically, we consider the following three allocation strategies for determining the set of blue nodes:
\begin{inparaenum}[(i)]
\item The {\em \topalg} strategy, which picks the set of $\numblue$ blue nodes as the set closest to the root.
This approach targets reducing the number of messages transmitted in the topmost part of the network, where congestion is expected to be largest.
\item The {\em \maxalg} strategy, which picks the set of blue nodes as the $\numblue$ switches with the largest 
load.
This approach is motivated by the fact that one should aim at reducing link congestion ``at the bud'', which would presumably have a positive effect on overall congestion.
\item The {\em \levelalg} strategy, defined for complete binary trees, which aims at partitioning the network into subtrees of similar size, where all the messages within a subtree are aggregated. This is done  by picking a whole level in the complete binary tree as the set of blue nodes.
This approach, which essentially targets load balancing, strives to ``equalize'' congestion in distinct sub-trees in the network.
\end{inparaenum}

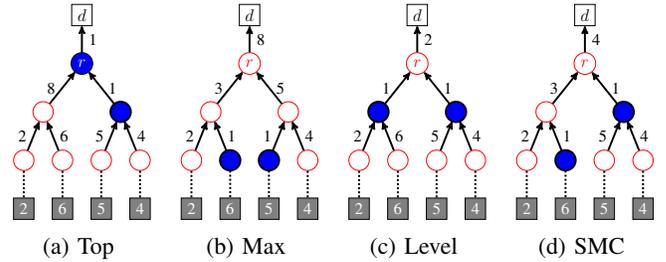
\begin{figure}[t]
    \centering
    \subcaptionbox{\topalg\label{fig:toy_example_1:top}}{
        \resizebox{0.21\columnwidth}{!}{\begin{tikzpicture}
\node
    [node_p] {$\destination$}
        child{ node [node_b] {$\rootswitch$} 
                child{  node [node_r] {} 
                        child{  node [node_r] (s1) {}
                                child{  node [node_h] (h1) {2}
                                        edge from parent[draw=none]
                                }
            				    edge from parent node[left,xshift=-2mm] {2}
                        }
                        child{  node [node_r] (s2) {}
                                child{  node [node_h] (h2) {6}
                                        edge from parent[draw=none]
                                }
            				    edge from parent node[right,xshift=2mm] {6}
                        }
                        edge from parent node[left,xshift=-2mm] {8}
                }
                child{  node [node_b] {} 
                        child{  node [node_r] (s3) {}
                                child{  node [node_h] (h3) {5}
                                        edge from parent[draw=none]
                                }
            				    edge from parent node[left,xshift=-2mm] {5}
                        }
                        child{  node [node_r] (s4) {}
                                child{  node [node_h] (h4) {4}
                                        edge from parent[draw=none]
                                }
            				    edge from parent node[right,xshift=2mm] {4}
                        }
                        edge from parent node[right,xshift=2mm] {1}
                }
                edge from parent node[right,xshift=2mm] {1}
        }
;

\path[-,dashed,line width=1mm,]
    (s1) edge (h1)
    (s2) edge (h2)
    (s3) edge (h3)
    (s4) edge (h4)
;
    
\end{tikzpicture}}
    }
    \subcaptionbox{\maxalg\label{fig:toy_example_1:max}}{
        \resizebox{0.21\columnwidth}{!}{\begin{tikzpicture}
\node
    [node_p] {$\destination$}
        child{ node [node_r] {$\rootswitch$} 
                child{  node [node_r] {} 
                        child{  node [node_r] (s1) {} 
                                child{  node [node_h] (h1) {2}
                                        edge from parent[draw=none]
                                }
            				    edge from parent node[left,xshift=-2mm] {2}
                        }
                        child{  node [node_b] (s2) {}
                                child{  node [node_h] (h2) {6}
                                        edge from parent[draw=none]
                                }
            				    edge from parent node[right,xshift=2mm] {1}
                        }
                        edge from parent node[left,xshift=-2mm] {3}
                }
                child{  node [node_r] {} 
                        child{  node [node_b] (s3) {}
                                child{  node [node_h] (h3) {5}
                                        edge from parent[draw=none]
                                }
            				    edge from parent node[left,xshift=-2mm] {1}
                        }
                        child{  node [node_r] (s4) {}
                                child{  node [node_h] (h4) {4}
                                        edge from parent[draw=none]
                                }
            				    edge from parent node[right,xshift=2mm] {4}
                        }
                        edge from parent node[right,xshift=2mm] {5}
                        }
                edge from parent node[right,xshift=2mm] {8}
        }
;

\path[-,dashed,line width=1mm,]
    (s1) edge (h1)
    (s2) edge (h2)
    (s3) edge (h3)
    (s4) edge (h4)
;

\end{tikzpicture}}
    }
    \subcaptionbox{\levelalg\label{fig:toy_example_1:level}}{
        \resizebox{0.21\columnwidth}{!}{\begin{tikzpicture}
\node
    [node_p] {$\destination$}
        child{ node [node_r] {$\rootswitch$} 
                child{  node [node_b] {} 
                        child{  node [node_r] (s1) {}
                                child{  node [node_h] (h1) {2}
                                        edge from parent[draw=none]
                                }
            				    edge from parent node[left,xshift=-2mm] {2}
                        }
                        child{  node [node_r] (s2) {}
                                child{  node [node_h] (h2) {6}
                                        edge from parent[draw=none]
                                }
            				    edge from parent node[right,xshift=2mm] {6}
                        }
                        edge from parent node[left,xshift=-2mm] {1}
                }
                child{  node [node_b] {} 
                        child{  node [node_r] (s3) {}
                                child{  node [node_h] (h3) {5}
                                        edge from parent[draw=none]
                                }
            				    edge from parent node[left,xshift=-2mm] {5}
                        }
                        child{  node [node_r] (s4) {}
                                child{  node [node_h] (h4) {4}
                                        edge from parent[draw=none]
                                }
            				    edge from parent node[right,xshift=2mm] {4}
                        }
                        edge from parent node[right,xshift=2mm] {1}                }
                edge from parent node[right,xshift=2mm] {2}
        }
;

\path[-,dashed,line width=1mm,]
    (s1) edge (h1)
    (s2) edge (h2)
    (s3) edge (h3)
    (s4) edge (h4)
;

\end{tikzpicture}}
    }
    \subcaptionbox{\alg\label{fig:toy_example_1:alg}}{
        \resizebox{0.21\columnwidth}{!}{\begin{tikzpicture}
\node
    [node_p] {$\destination$}
        child{ node [node_r] {$\rootswitch$} 
                child{  node [node_r] {} 
                        child{  node [node_r] (s1) {} 
                                child{  node [node_h] (h1) {2}
                                        edge from parent[draw=none]
                                }
            				    edge from parent node[left,xshift=-2mm] {2}
                        }
                        child{  node [node_b] (s2) {}
                                child{  node [node_h] (h2) {6}
                                        edge from parent[draw=none]
                                }
            				    edge from parent node[right,xshift=2mm] {1}
                        }
                        edge from parent node[left,xshift=-2mm] {3}
                }
                child{  node [node_b] {} 
                        child{  node [node_r] (s3) {} 
                                child{  node [node_h] (h3) {5}
                                        edge from parent[draw=none]
                                }
            				    edge from parent node[left,xshift=-2mm] {5}
                        }
                        child{  node [node_r] (s4) {}
                                child{  node [node_h] (h4) {4}
                                        edge from parent[draw=none]
                                }
            				    edge from parent node[right,xshift=2mm] {4}
                        }
                        edge from parent node[right,xshift=2mm] {1}
                }
                edge from parent node[right,xshift=2mm] {4}
        }
;

\path[-,dashed,line width=1mm,]
    (s1) edge (h1)
    (s2) edge (h2)
    (s3) edge (h3)
    (s4) edge (h4)
;
\end{tikzpicture}}
    }
    \caption{Example of solutions produced by 4 allocation algorithms for a simple load over a weighted tree network, with constant rates of $1$ and $\numblue=2$ aggregation switches (blue nodes).
    }
    \label{fig:toy_example_1}
\end{figure}

Consider a tree network with $n=7$ switches which induces a complete binary tree topology on the set of switches which are all available for aggregation, with a constant rate of $1$ for all links.
Servers are connected only to leaf switches.
Such a topology can be viewed as if the leaf switches are effectively top-of-rack (ToR) switches in a small datacenter topology, where each rack accommodates a distinct number of servers (or VMs).
Fig.~\ref{fig:toy_example_1} provides an illustration of the network.
Each leaf switch is connected to a rack of several worker servers where the number of workers in the rack is marked in the gray square. 
In particular, the load handled by the 4 leaf switches is $(2,6,5,5)$ (from left to right).
In our example the maximum number of blue switches allowed is set to $\numblue=2$.
Each link $e$ is marked with its link congestion, $\util_e(\network,\load,\blueset)$. 

Figs.~\eqref{fig:toy_example_1:top}, \eqref{fig:toy_example_1:max}, and~\eqref{fig:toy_example_1:level} show the results of applying strategies \topalg, \maxalg, and \levelalg, respectively, to such a network and load, obtaining a network congestion of 8, 9, and 6, respectively.
The optimal approach, which is obtained by our proposed algorithm, \alg\ (formally described and analyzed in Sec.~\ref{sec:algorithm}), ends up picking a non-trivial set of blue nodes, as can be seen in Fig.~\eqref{fig:toy_example_1:alg}.
This allocation strictly outperforms all three contending strategies, a network congestion of 5.
As we show in the sequel, our algorithm is optimal, and thus ensures to have the minimum congestion possible.

A further observation, which hinders the applicability of greedy approaches, is that the optimal solution is not necessarily monotone in $\numblue$.
For the network in Fig.~\ref{fig:toy_example_1}, one may consider the optimal placement for $\numblue=2,3,4$.. There is no way to add a single blue node to the optimal solution for $k=2$ and obtain an optimal set of blue nodes for $k=3$, that is a subset of the optimal solution for $k=4$.


\section{\alg: An Optimal Algorithm}
\label{sec:algorithm}

In this section we describe our algorithm, Search for Minimal Congestion (\alg), that produces an optimal solution to the \combic\ problem.
 The main technical contribution of the paper is the following theorem.
\begin{theorem}
\label{thm:alg_is_optimal}
Given a weighted tree network $\network$ with rates $\weight$, a load $\load$, availability $\Availability$, and a bound $\numblue$ on the number of allowed blue switches, algorithm \alg\ solves the \combic\ problem in time $O\left( n\cdot \numblue^2\cdot \log \left( \frac{\weight_{\max}}{\weight_{\min}} \cdot \sum_{v}\load(v) \right) \right)$.
\end{theorem}

\subsection{Overview of \alg}

\algblockdefx[RUN]{StartRun}{EndRun}[1]{{\bf run} #1}{}
\algnotext[RUN]{EndRun}

\begin{algorithm}[t!]
\caption[Algorithm]{\alg$(\network,\load,\Availability,\numblue)$}
\label{alg:smc}
\begin{algorithmic}[1]
\Require A tree $T$, load $\load$, availability $\Availability$, $\numblue$ blue nodes
\State $\mbn=\frac{1}{\min_e \weight(e)}\sum_v \load(v)$
\Comment{init. congestion upper bound}
\State $S = \frac{1}{\max_e \weight(e)}$
\StartRun {binary search in the range $[0,X]$ with step size $S$, using \alggather, finding the {\em minimal} congestion upper bound $\mbn^*$, returning the corresponding $\up^*$}
    
\EndRun
\StartRun{\algcolor$(\numblue)$ using $\up^*$}
\EndRun
\end{algorithmic}
\end{algorithm}

In this section we provide a bird's-eye view of \alg, which is formally defined in Algorithm~\ref{alg:smc}.
The algorithm runs a binary search for the minimal congestion for which a feasible solution $\blueset \subseteq \Availability$ exists.
Given the bound $\numblue$ on the number of blue nodes allowed in the network, for each potential upper bound $X$ on the congestion, \alg\ uses dynamic programming, and is split into two phases.

The algorithm used during the binary search in the first phase, dubbed \alggather, consists of scanning the switches in the tree in DFS-order.
In every switch node $v$ we effectively consider all {\em potentially efficient partitions} of any number $i \leq \numblue$ of blue nodes across all children of the node.
For every such $i$, the partition that minimizes the number of messages leaving the node is retained (maintained by the vector $\up_v$), and information is passed on to the parent of the node.
We note that the algorithm finds such a partition efficiently. 
The main property satisfied by \alggather\ is shown in Lemma~\ref{lem:gather_correctness:induction}.
The information disseminated upwards by \alggather\ is then used in the second phase to compute the optimal solution (and place the blue nodes).
\alggather\ is formally defined in Algorithm~\ref{alg:alg:gather}, where it is described as an asynchronous distributed algorithm, with synchronization induced by messages sent from a node to its parent.

In the second phase we apply algorithm \algcolor, which scans the nodes of the tree in reverse-DFS-order, and essentially tracks the feasible allocation satisfying the upper bound $X$ on the congestion (if such an allocation exists).
Initially a node is considered red, and during the scan \algcolor\ sets a node as blue only when it is necessary for satisfying the congestion constraint determined by the upper bound $X$ (if possible).
A node then informs each of its children as to the number of (remaining) blue nodes that can be distributed in the subtree rooted at that child.
To this end, \algcolor\ uses the information obtained by \alggather, and in particular the partition that ensures that the congestion constraint is satisfied (if possible).
\algcolor\ is formally defined in Algorithm~\ref{alg:alg:color}, where it is also described as an asynchronous distributed algorithm. Here synchronization is induced by messages received by a node from its parent.



\subsection{Analysis of \alg}
\label{sec:analysis}


We begin by introducing some notation that would be used throughout our proofs.
For very node $v$, we let $c_1,\ldots,c_{C(v)}$ denote the children of $v$ (in some arbitrary fixed order).
For every $m=1,\ldots,\childnum(v)$ we let $\network_v^m$ denote the subtree rooted at $v$ containing only the subtrees rooted at children $c_1,\ldots,c_m$, and let $\tilde{\network}_v^m$ denote the {\em extended subtree} of $\network_v^m$, which is extended by adding the link $(v,\parent(v))$.
We further let $\network_v=\network_v^{\childnum(v)}$ denote the subtree rooted at $v$ (containing all subtrees of all children of $v$), and let $\tilde{\network}_v$ be the extended subtree of $\network_v$.

Let $\mbn$ be a real value, representing an upper bound on {\em network congestion}.
We define $\up(\network_v,\load,\numblue,\mbn)$ as the minimum number of messages traversing link $(v,\parent(v))$ for which there exists a set $\blueset\subseteq\network_v,\abs{\blueset}=\numblue$ that satisfies the   congestion constraint $\msgcost(\tilde{\network_v},\load,\blueset)\leq \mbn$ (or infinity if no such set exists).%
\footnote{Note that the congestion constraint should be satisfied also for link $(v,\parent(v))$.}



Given some value $\mbn$, algorithm \alggather\ uses the following concepts for non-leaf nodes:
\begin{inparaenum}[(i)]
\item variables $\up_v^m(i,\nodecolor)$ that should represent the minimum number of messages traversing link $(v,\parent(v))$ in the tree $\tilde{\network}_v^m$, where $v$ is colored by $\nodecolor$ and at most $i$ nodes in $\network_v^m$ are blue, while ensuring that the congestion in $\tilde{\network}_v^m$ is at most $\mbn$, and
\item variables $\up_v(i)=\min \set{\up_v^{C(v)}(i,\blue),\up_v^{C(v)}(i,\red)}$.
\end{inparaenum}
In the following lemma we prove that the semantics we attribute to $\up_v^m(i,\nodecolor)$ are indeed correct, and that \alggather\ indeed computes $\up(\network_v,\load,i,\mbn)$ correctly.

\begin{lemma}
\label{lem:gather_correctness:induction}
For every node $v$, every $m=1,\ldots,\childnum(v)$, and every $i=0,\ldots,k$,
$\up_v(i)$ as computed by \alggather\ satisfies $\up_v(i)=\up(\network_v,\load,i,\mbn)$, where
if $v$ is not a leaf then $\up_v^m$ as computed by \alggather$(\network,\load,\Availability,\numblue,\mbn)$ satisfies

\begin{align}
\up_v^m(i,\red) &= \up(\network_v^m,\load,i,\mbn) \mbox{ where $v$ is colored $\red$}
\label{lem:eq:beta_v_m_i_red}
\end{align}
and
\begin{align}
\up_v^m(i,\blue) &= \up(\network_v^m,\load,i,\mbn) \mbox{ where $v$ is colored $\blue$},
\label{lem:eq:beta_v_m_i_blue}
\end{align}
where
\begin{align}
\up_v^1(i,\blue)
&= \begin{cases}
1, & \text{if } \up_{c_1}(i-1) < \infty \\
\infty, & \text{otherwise},
\end{cases}
\label{lem:eq:first_child_blue}\\
\up_v^1(i,\red)&= \begin{cases}
 \up_{c_1}(i)+\load(v), & \text{if } (\up_{c_1}(i)+\load(v))\cdot \rate(v) \leq \mbn\\
\infty, & \text{otherwise}
\end{cases}
\label{lem:eq:first_child_red}
\end{align}
and for $m>1$
\begin{align}
\up_v^m(i,\blue)
&= \left\{
\begin{array}{ll}
\!\!\!1, & \!\!\!\text{if } \displaystyle{\min_{0 \le j < i}}(\up_v^{m-1}(i-1-j,\blue) + \up_{c_m}(j)) < \infty \\
\!\!\!\infty, & \!\!\!\text{otherwise}
\end{array}
\right.
\label{lem:eq:m_child_blue}\\
\up_v^m(i,\red)
&= \begin{cases}
\displaystyle{\min_{ 0 \le j \le i}(\up_v^{m-1}(i-j,\red) + \up_{c_m}(j))}, & \text{if \eqref{lem:eq:congestion_condition} holds} \\
\infty, & \text{otherwise}
\end{cases}
\label{lem:eq:m_child_red}
\end{align}
where
\begin{align}
  \displaystyle{\min_{ 0 \le j \le i }(\up_v^{m-1}(i-j,\red) + \up_{c_m}(j))}\cdot \rate(v) \leq \mbn. \label{lem:eq:congestion_condition} 
\end{align}
Overall,
\begin{align}
\up_v(i)
&= \min \left( \up_v^{C(v)}(i,\blue),\up_v^{C(v)}(i,\red) \right)=\up(\network_v,\load,i,\mbn)
\label{lem:eq:beta_v_i}
\end{align}

\end{lemma}

\revision{
\begin{IEEEproof}[Proof]
The proof is by double induction on the height of $\network_v$ and the number of children $m$ for which 
$\up_v^m(i,\red)$ and $\up_v^m(i,\blue)$ have been computed correctly.

For the base case, we observe that for any leaf node $v$ the following holds:
\begin{inparaenum}[(i)]
    \item For $i>0$, $v$ can be colored blue, and this minimizes the load on link $(v,\parent(v))$ implying that $\up_v(i)=1$.
    \item For $i=0$, $v$ cannot be colored blue, the load on the outgoing link is $\load(v) \cdot \rate(v,\parent(v))$, implying that:
    \begin{align}
        \up_v(0)&=\begin{cases}
        \load(v), & \text{if } \load(v) \cdot \rate(v,\parent(v)) \leq \mbn\\
        \infty, & \text{otherwise.}
        \end{cases}
    \end{align}
\end{inparaenum}
It follows that for every leaf node $v$,
\begin{align}
    \up_v(i)=\up(\network_v,\load,i,\mbn),
    \label{eq:induction_base}
\end{align}
which proves the base case.

Let $v$ be a non leaf and assume that $\up_{v'}(i)$ has been computed correctly for all nodes $v'$ at height less then node $v$'s height, and for all $i$. In particular, this is true for every child $c_m$ of node $v$, $m=1,\ldots,\childnum(v)$.
Consider first $m=1$, where we have two cases:
\begin{enumerate}[(i)]
\item Assume $v$ is blue and $i>0$.
By the induction hypothesis, if $\up_{c_1}(i-1)< \infty$, i.e. satisfies the congestion constraint, then $\up_v^1(i,\blue)=1$. Otherwise, again by the induction hypothesis, if $\up_{c_1}(i-1)=\infty$ both $\up_{c_1}(i-1)$ and $\up_v^1(i,\blue)$ don't satisfy the congestion constraint.
Eq.~\ref{lem:eq:first_child_blue} follows.

\item Assume $v$ is red. By the induction hypothesis, if $\up_{c_1}(i)< \infty$, then there is a solution that satisfies the congestion constraint using $i$ blue nodes in $\tilde{\network}_{c_1}$. 
If $(\up_{c_1}(i)+\load(v)) \cdot \rate(v) \leq \mbn$ then the congestion constraint is also satisfied on $(v,\parent(v))$ in $\tilde{\network}_v^1$,
implying that $\up_v^1(i,\red)=\up_{c_1}(i)+\load(v)$.
Otherwise the congestion constraint is violated either in 
Eq.~\ref{lem:eq:first_child_red} follows.
\end{enumerate}

Now consider $m>1$,
where we assume that for all $m'<m$, $\up_v^{m'}(i,\red)$ and $\up_v^{m'}(i,\blue)$ have been computed correctly, and in particular, satisfy Eq.~\ref{lem:eq:beta_v_m_i_red} and~\ref{lem:eq:beta_v_m_i_blue}.
We distinguish between two cases:
\begin{enumerate}[(i)]

\item Assume $v$ is blue and $i>0$.
If there exists a $j$ such that, $\up_{c_m}(j) < \infty$ 
and $\up_v^{m-1}(i-1-j) < \infty$, by the induction hypothesis, this means that the congestion constraint is satisfied both in $\tilde{\network}_{c_m}$ with $j$ blue nodes and $\network_{v}^{m-1}$ with $i-1-j$ blue nodes.

This implies that there exists a partition of $i$ that satisfies the congestion constraint, and $\up_v^{m}(i,\blue)=1$.
Otherwise, the congestion constraint cannot be satisfied by {\em any} partition, in which case $\up_v^{m}(i,\blue)=\infty$.
Eq.~\ref{lem:eq:m_child_blue} and~\ref{lem:eq:beta_v_m_i_blue} thus follow.

\item Assume $v$ is red, and that there exists a $j$ such that, $\up_{c_m}(j) < \infty$ 
and $\up_v^{m-1}(i-j) < \infty$.
For each such $j$, by the induction hypothesis, the congestion constraint is satisfied by this partition both in $\tilde{\network}_{c_m}$ with $j$ blue nodes and $\network_{v}^{m-1}$ with $i-j$ blue nodes.
If, additionally, $(\up^{(m-1)}_v(i-j)+\up_{c_m}(j)) \cdot \rate(v) \leq \mbn$ then the congestion constraint is also satisfied on $(v,\parent(v))$ by this partition.
Taking the minimum over all such partitions ensures that the number of messages traversing $(v,\parent(v))$ is minimized, while satisfying the congestion constraint in $\tilde{\network}_v^m$.
To see this, assume by contradiction that there exists a way to have less messages traverse $(v,\parent(v))$ while satisfying the congestion constraint.
In particular, such a solution places some $j$ blue nodes in $\network_{c_m}$, and $(i-j)$ blue nodes in $\network_v^{m-1}$.
Since the additional load on $(v,\parent(v))$ due to $\load(v)$ is independent of any such placement, it follows that having a smaller number of messages traverse $(v,\parent(v))$ implies that either the number of messages traversing $(c_m,v)$ is smaller than $\up_{c_m}(j)$ or smaller than $\up(\network_v^{m-1},\load,i,\mbn)$, contradicting the correctness of $\up_{c_m}(j)$ or $\up_v^m(i-j,\red)$, respectively, which follows from the induction hypothesis.
This shows the validity of Eq.~\ref{lem:eq:m_child_red} and~\ref{lem:eq:beta_v_m_i_red}, which completes the proof.


\end{enumerate}
\end{IEEEproof} }

\usetikzlibrary{fit,calc}
\newcommand*{\tikzmk}[1]{\tikz[remember picture,overlay,] \node (#1) {};\ignorespaces}
\newcommand{\boxit}[5]{\tikz[remember picture,overlay]{\node[yshift=3pt,fill=#1,opacity=.25,fit={($(A)-(#2\linewidth,#3\baselineskip)$)($(B)+(#4\linewidth,#5\baselineskip)$)}] {};}\ignorespaces}
%
\colorlet{mode1}{red!40}
\colorlet{mode2}{cyan!60}
\colorlet{mode3}{green!70}
\colorlet{mode4}{gray!70}

\makeatletter
\newcommand\footnoteref[1]{\protected@xdef\@thefnmark{\ref{#1}}\@footnotemark}
\makeatother

\begin{algorithm}[t!]
\caption[Algorithm]{\alggather$(\network,\load,\Availability,\numblue,\mbn)$ at node $v$}
\label{alg:alg:gather}
\begin{algorithmic}[1]
\Require A tree $T$, load $\load$, availability $\Availability$, $\numblue$ $\#$ of blue nodes and $\mbn$ maximal link utilization.
\Ensure Correct potential functions, $\up_v$, at each node $v$
\If{$v$ is a leaf node}
    \tikzmk{A}
            \State $\up_v(0)=\load(v) $
        \If{$ \up_v(0) \cdot \rate(v,\parent(v)) > \mbn$}
            \State $\up_v(0)=\infty$
        
        \EndIf
        \label{alg:gather:leaf:i_equals_0}
        \For{$i=1,\dots,\numblue$}
        \Comment{$v$ can be blue}
            \If{$v \in \Availability$}
            \Comment{$v$ is available}
            
                \State $\up_v(i)= 1 $
            \Else
                \State $\up_v(i)= \up_v(0) $
            \label{alg:gather:leaf:i_greater_0}
            \EndIf
        \EndFor
    \State send $\up_v$ to $\parent(v)$ and \Return
    \Comment{inform parent}
\EndIf
\tikzmk{B}
\boxit{mode1}{0.475}{-0.15}{0.987}{-0.1}

\State wait to receive $\up_c$ from each child $c$ of $v$
\For{$m=1,\ldots,C(v)$}
    \label{alg:for:ell:start}
    \State $c_m \gets$ the $m$'th child of $v$
        \For{$i=0,\dots,\numblue$}
            \label{alg:for:num_blue}
            \If{$m=1$}
                \tikzmk{A}
                \State $\up_v^m(i,R)= \up_{c_m}(i)+ \load(v) $
                \label{alggather:line:first_child_red_start}
                \If{$\up_v^m(i,R)\cdot \rate(v,\parent(v)) > \mbn$}
                    \State $\up_v^m(i,R)=\infty$
                \label{alggather:line:first_child_red_end}
                \EndIf
                \If{$i > 0 \AND \up_{c_m}(i-1)\leq \mbn \AND v \in \Availability$}
                \label{alggather:line:first_child_blue_start}
                    \State $\up_v^m(i,B)= 1$  
                \Else
                    \State $\up_v^m(i,B)= \infty$ 
                    \label{alggather:line:first_child_blue_end}
                \EndIf
                \label{alg:gather:Y_v^m:first_child:red}
                \tikzmk{B}
                \boxit{mode2}{0.4}{-0.1}{0.86}{0.05}
            \Else
            \Comment{$m > 1$}
                \tikzmk{A}
                \State $\up_v^m(i,B)= \mincost(i-1,\up_v^{m-1},\up_{c_m},\mbn,\blue)$                \footnote{\label{note1}When $i=0$ then $\up_v^m(i,\blue)= \infty$.}

                \label{alggather:line:m_child_blue}
                \State $\up_v^m(i,R)= \mincost(i,\up_v^{m-1},\up_{c_m},\mbn,\red)$
                \label{alggather:line:m_child_red}
                \tikzmk{B}
                \boxit{mode3}{1.05}{-0.05}{0.145}{-0.15}
            \EndIf
    \EndFor
\EndFor
    \For{$i=0,\ldots,\numblue$}
        \State $\up_v(i)=\min \set{\up_v^{C(v)}(i,\blue),\up_v^{C(v)}(i,\red)}$
        \label{alggather:line:beta_v_i}        

    \EndFor
\State send $\up_v$ to $\parent(v)$ and \Return
\Statex \hrulefill
\Procedure{$\mincost(i,\up_v^{m-1},\up_{c_m},\mbn,\nodecolor)$}{}
\label{alg:mincost:start}
    \State $\displaystyle{\up=\min_{0\leq j \leq i }[\up_v^{m-1}(i-j,\nodecolor)+\up_{c_m}(j)]}$
    \If{$\up \cdot \rate(v,\parent(v))> \mbn$}
        \State \Return $\infty$
    \Else
        \State \Return $\up$
    \EndIf


\EndProcedure
\label{alg:mincost:end}
\end{algorithmic}
\end{algorithm}

\begin{algorithm}[t!]
\caption[Algorithm]{\algcolor$(k)$ at node $v$}
\label{alg:alg:color}
\begin{algorithmic}[1]
\Require $\up$
\Ensure Optimal coloring
\If{$v$ is the destination $d$}
    \State send $k$ to $r$ and \Return \label{alg:algcolor:dest}
\EndIf
\State color $v$ red and wait for $i$ from $\parent(v)$
\Statex \Comment{$i$: number of blue nodes in $\network_v$} 
\If{$v$ is a leaf node and $i>0$}
\tikzmk{A}
    \State color $v$ blue and \Return
    \label{alg:algcolor:condition_for_blue_leaf}
\EndIf
\tikzmk{B}
\boxit{mode1}{0.665}{-0.15}{0.985}{-0.05}
\If{$\up^{C(v)}_v(i,\blue) < \infty$}
    \Comment{$\up^{C(v)}_v(i,\blue) < \up^{C(v)}_v(i,\red)$}
    \label{alg:algcolor:condition_for_blue}
    \State color $v$ blue
\EndIf
\For{$m=C(v),\ldots,2$}
\Comment{children in reverse order}
\tikzmk{A}
    \State $j=\minsplit(i,\up_v^{m-1},\up_{c_m},\mbox{color of }v)$
    \label{alg:algcolor:msplit}
    \State send $j$ to $c_m$
    \State $i = i-j$
\EndFor
\tikzmk{B}
\boxit{mode3}{1.054}{-0.15}{0.978}{-0.18}
\If{$v$ is blue}
\Comment{handle $c_1$ last}
\tikzmk{A}
    \State send $i-1$ to $c_1$
\Else
    \State send $i$ to $c_1$
\EndIf
\tikzmk{B}
\boxit{mode2}{1.054}{-0.15}{0.978}{-0.05}
\State \Return
\Statex \hrulefill
\Procedure{$\minsplit(i,\up_v^{m-1},\up_{c_m},\nodecolor)$}{}
\label{alg:minsplit:start}
    \If{$\nodecolor==\red$}
        \State \Return $\displaystyle{\argmin_{0\leq j \leq i }[\up_v^{m-1}(i-j,\nodecolor)+\up_{c_m}(j)]}$
    \Else
    \Comment{$\nodecolor==\blue$}
        \State \Return $\displaystyle{\argmin_{0\leq j < i }[\up_v^{m-1}(i-j,\nodecolor)+\up_{c_m}(j)]}$
        
    \EndIf
\EndProcedure
\label{alg:minsplit:end}
\end{algorithmic}
\end{algorithm}

In the second phase of \alg, \algcolor\ essentially traces back the allocation of blue nodes along the optimal path in the dynamic programming performed by \alggather.
To show that \algcolor\ indeed produces an optimal solution to the \combic\ problem we make use of the following lemma.

\begin{lemma}
\label{lem:color_minimizes_outgoing_load}
Assume $\beta$ is the output of \alggather\ for the network congestion upper bound $\mbn$, such that $\up_r(k)$ is finite.
Then, \algcolor\ colors blue a set $\blueset$, such that $\abs{\blueset} \leq k$, and $\msgcost(\network,\load,\blueset) \leq \mbn$.
\end{lemma}
\revision{
\begin{IEEEproof}
In what follows, we say a node $v$ is \emph{correctly assigned} if: (i) it is colored so as to satisfy with the congestion constraint of the system,
(ii) it is allotted the number of blue nodes for $\network_v$ so as to satisfy with the congestion constraint of the system.
We prove by induction on the order of handling nodes by \algcolor\ that if node $v$ is correctly assigned 
then each of its children $c_m$, $m=1,\ldots,\childnum(v)$ is correctly assigned.

For the base case, consider node $\destination$, which should have $\numblue$ blue nodes in its subtree, it's color is trivially not blue (since $\destination$ is a server). 
So $\destination$ is correctly assigned. $\destination$ has a single child, $\rootswitch$, and by line~\ref{alg:algcolor:condition_for_blue} of \algcolor, along with Eq.~\eqref{lem:eq:beta_v_m_i_red} and Eq.~\eqref{lem:eq:beta_v_m_i_blue} $\rootswitch$ is colored correctly, since by line~\ref{alggather:line:beta_v_i} 
of \alggather, its color is the one satisfying congestion constraint. Clearly by line \ref{alg:algcolor:dest} in \algcolor\ $r$ is correctly.

Assume the claim holds for all nodes handled before node $v$, and consider node $v$ which is correctly assigned.
First, since $v$ is correctly colored,
by induction on the number of children of $v$ from $\childnum(v)$ to 1,
it is easy to show
that each child $c$ is assigned the correct number of blue nodes to be distributed in its subtree $\network_c$.
This follows from the fact that the $\minsplit$ procedure in lines~\ref{alg:minsplit:start}-\ref{alg:minsplit:end} of \algcolor\ essentially extract the value $j$ obtaining the minimum considered also by the $\mincost$ procedure in lines~\ref{alg:mincost:start}-\ref{alg:mincost:end} in \alggather.
Since each child $c$ is assigned correctly the correct number of blue nodes, by Lemma \ref{lem:gather_correctness:induction} and
line~\ref{alggather:line:beta_v_i} of \alggather, $c$ will be also colored correctly.
\end{IEEEproof}
}

We now show that the \combic\ problem can be reduced to computing $\up(\network,\load,\numblue,\mbn)$.

\begin{lemma}
\label{lem:search}
If $\up(\network,\load,\numblue,\mbn)$ can be computed in $\alpha$ time, then \combic$(\network,\load,\Availability,\numblue)$ is solved in time $\alpha \cdot \log(\sum_{v}\load(v) \cdot \frac{\weight_{\max}}{\weight_{\min}})$.
\end{lemma}
\begin{IEEEproof}
The proof follows directly from applying a binary search over the upper bound $\mbn$ on the network congestion, where the maximum such value is no larger than $\frac{1}{\weight_{\max}} \cdot \sum_v \load(v)$, and the granularity is at least $\frac{1}{\weight_{\min}}$, where in each iteration we check whether or not $\up(\network,\load,\numblue,\mbn)$ is finite, using Algorithm \alggather.
\end{IEEEproof}

We can now prove  Theorem~\ref{thm:alg_is_optimal}.
\begin{IEEEproof}[Proof of Theorem~\ref{thm:alg_is_optimal}]
The correctness of the algorithm follows from Lemmas \ref{lem:gather_correctness:induction}-\ref{lem:search}.
For the running time of \alg, we note that it is dominated by the running time of \alggather, which, in turn, is dominated by the for-loops in lines~\ref{alg:for:ell:start}-\ref{alggather:line:m_child_red}.
This loop handles every edge $(v,p(v))$ once, and for each edge the running time is $O(\numblue^2)$, resulting in a total running time for \alggather\ of $O(n\cdot \numblue^2)$.
By Lemma~\ref{lem:search}, performing the binary search requires running \alggather\ $O\left( \log(\sum_{v}\load(v) \cdot \frac{\weight_{\max}}{\weight_{\min}}) \right)$ times, resulting is a total running time for solving the \combic\ problem of $O\left( n\cdot \numblue^2\cdot \log \left( \frac{\weight_{\max}}{\weight_{\min}} \cdot \sum_{v}\load(v) \right) \right)$.
\end{IEEEproof}

\begin{figure*}
    \centering
    \subcaptionbox{constant ($\weight=1$) \label{fig:gain_constant}}{
            \includegraphics[width=0.31\textwidth]{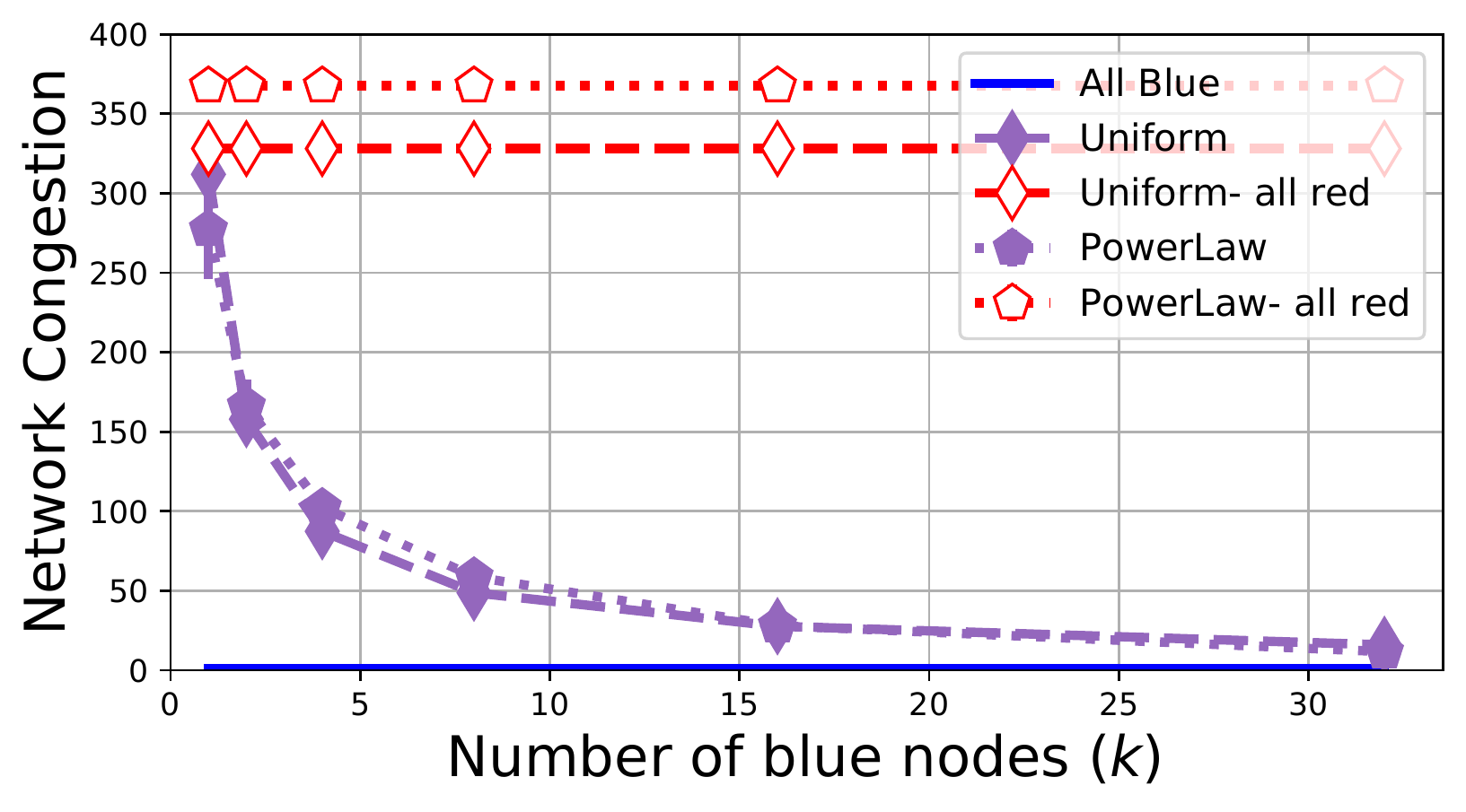} }
    \subcaptionbox{linear increasing ($\weight=i$) \label{fig:gain_linear}}{
            \includegraphics[width=0.31\textwidth]{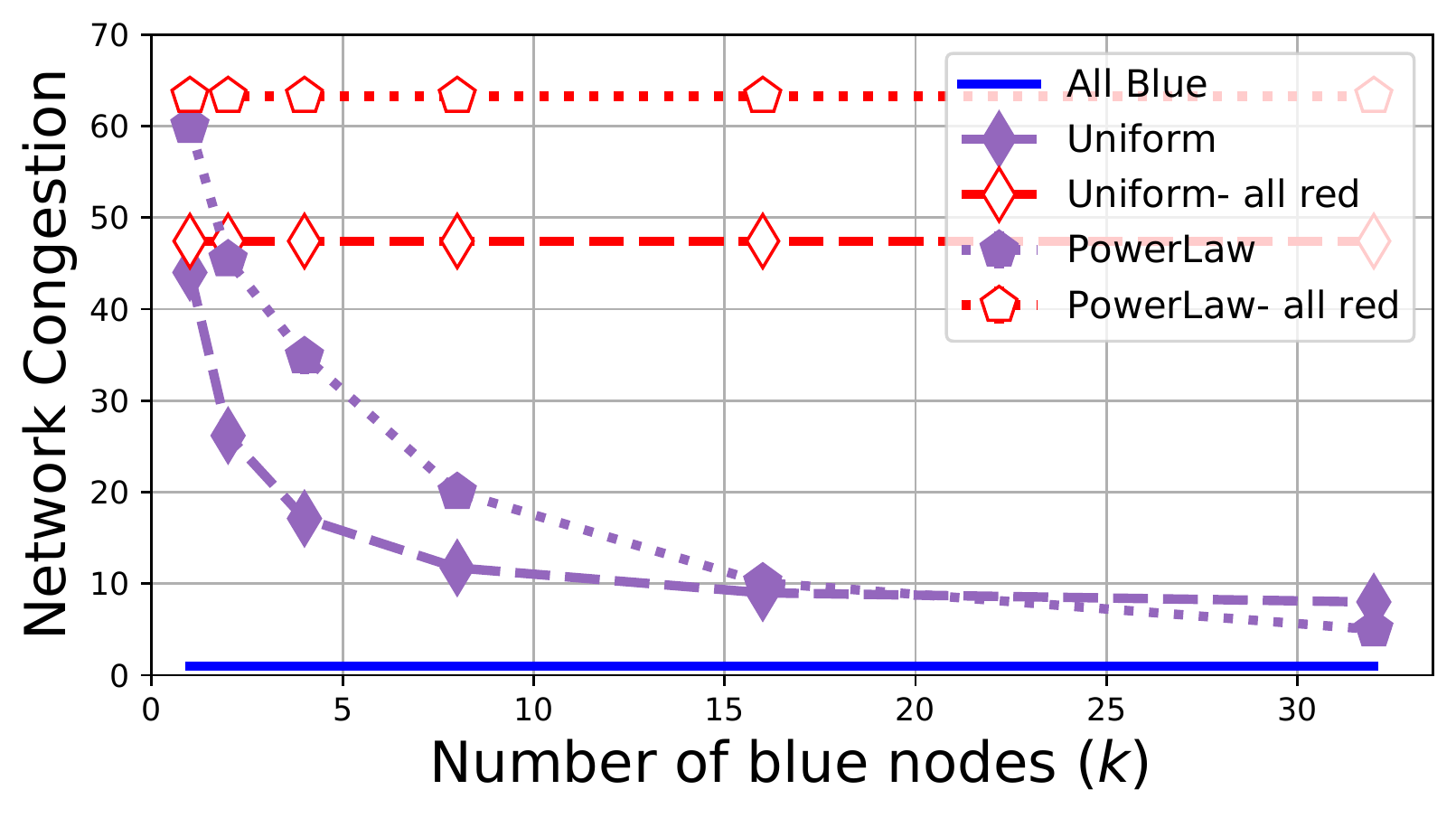} }
    \subcaptionbox{exponentially increasing ($\weight=(1.5)^i$) \label{fig:gain_exponential}}{
            \includegraphics[width=0.31\textwidth]{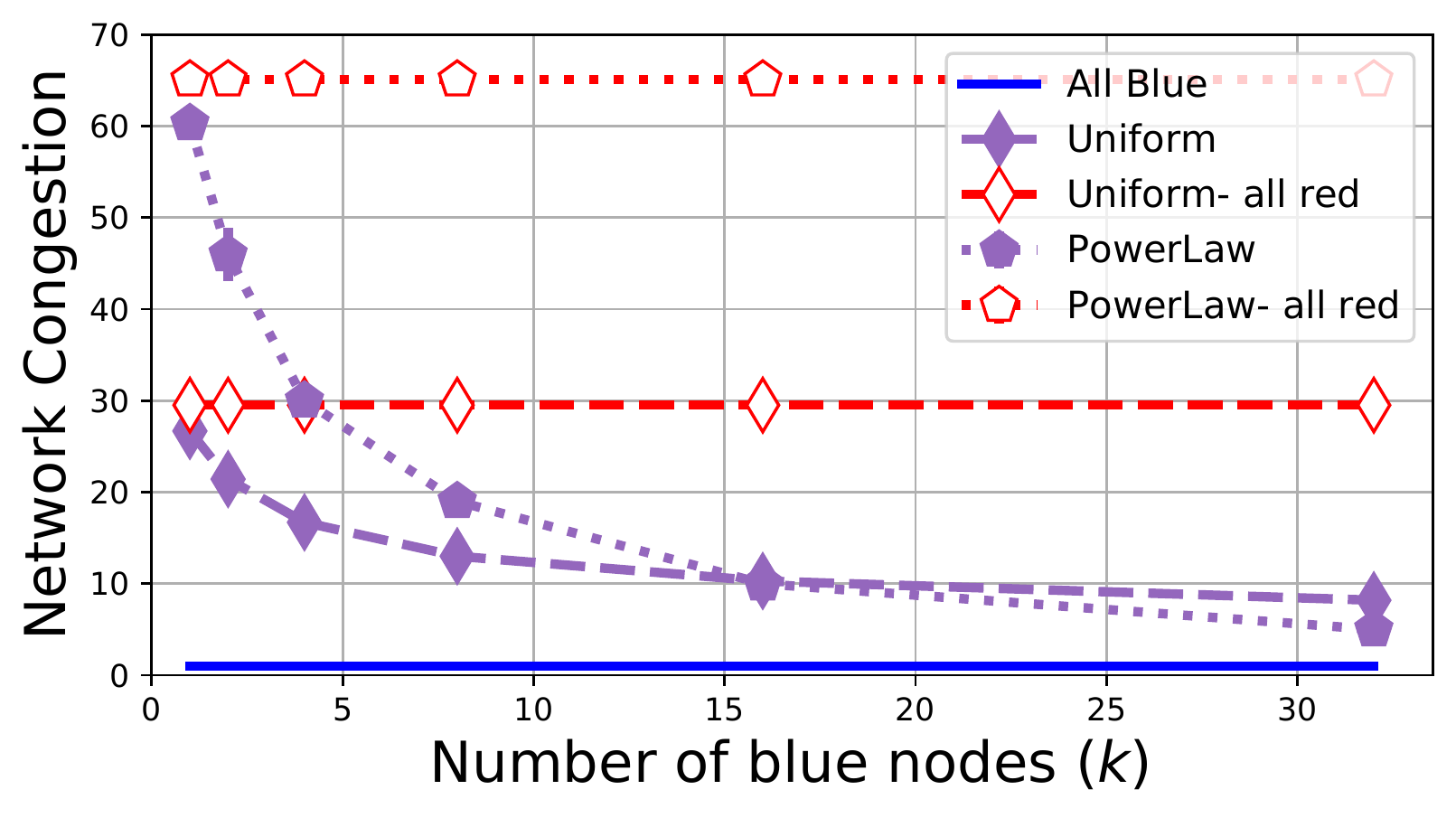} }
\caption{Limited In-network aggregation, \alg\ congestion gains with limited resources }
\label{fig:gains}
\end{figure*}

\begin{figure*}
    \centering
    \begin{tabular}{cccc}
        \raisebox{.25cm}{\rotatebox[origin=lc]{90}{\small Power-law load dist.}} \hspace{0.1cm} &
            \includegraphics[width=0.29\textwidth]{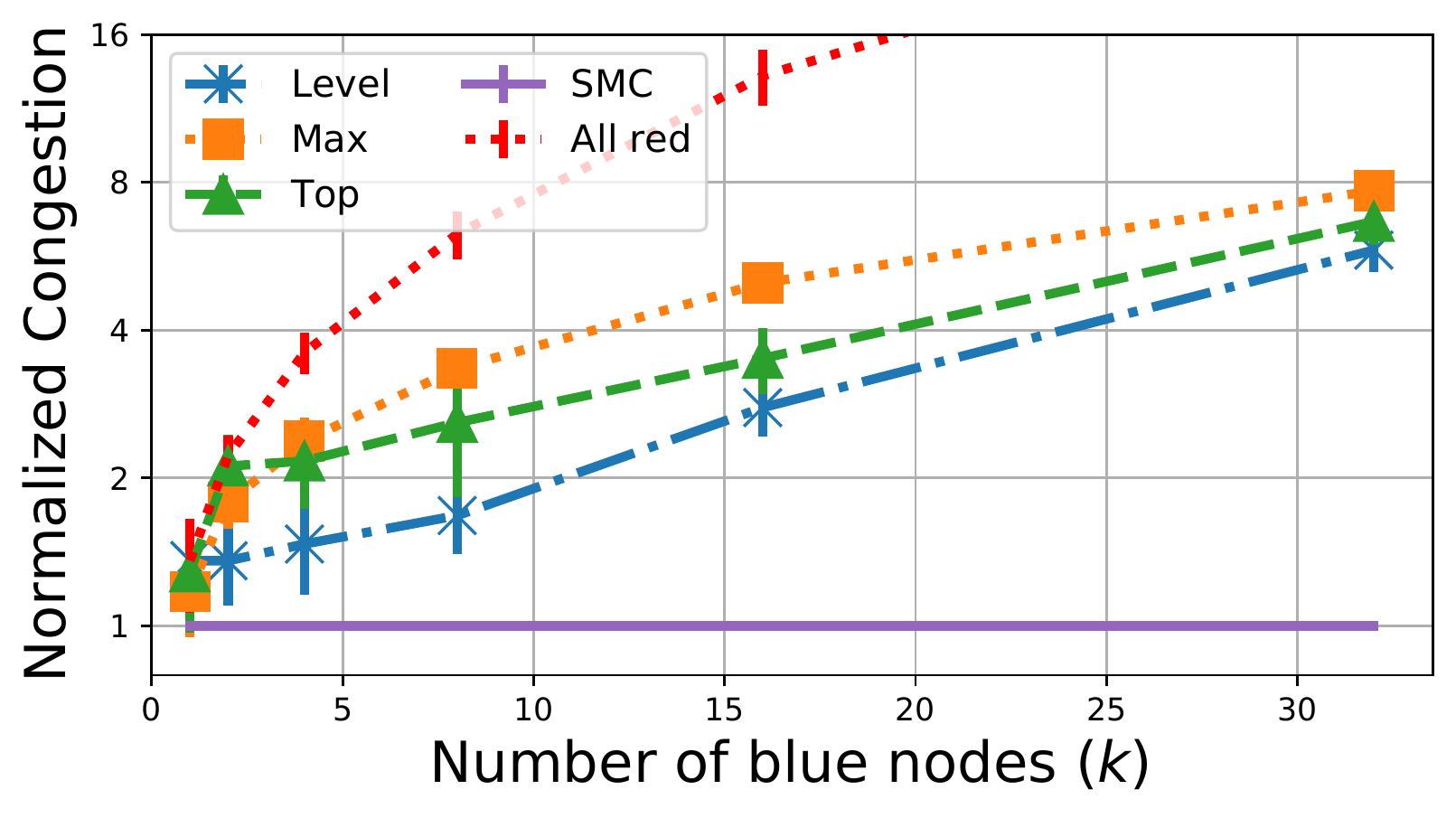} &
            \includegraphics[width=0.29\textwidth]{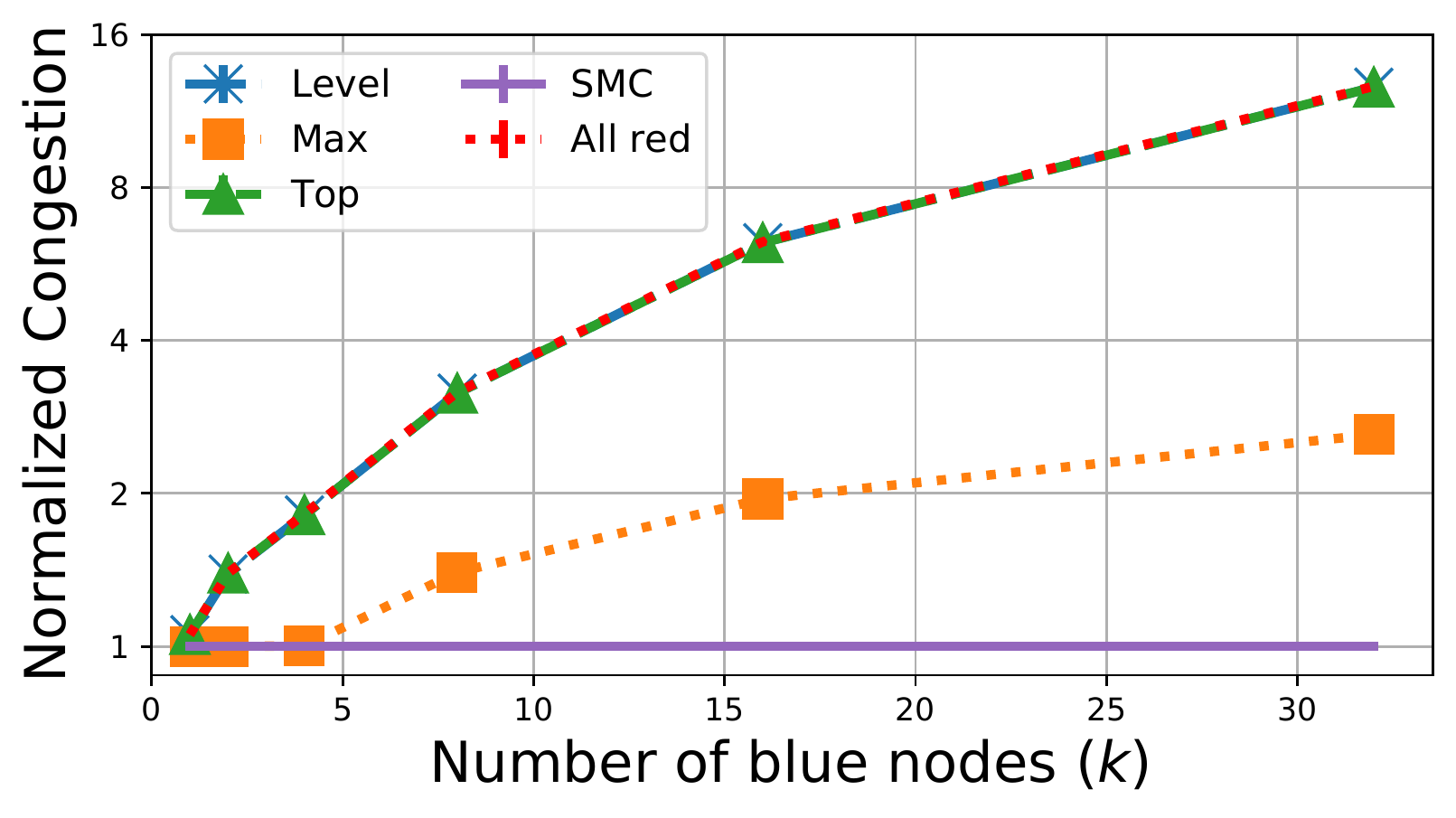} &
            \includegraphics[width=0.29\textwidth]{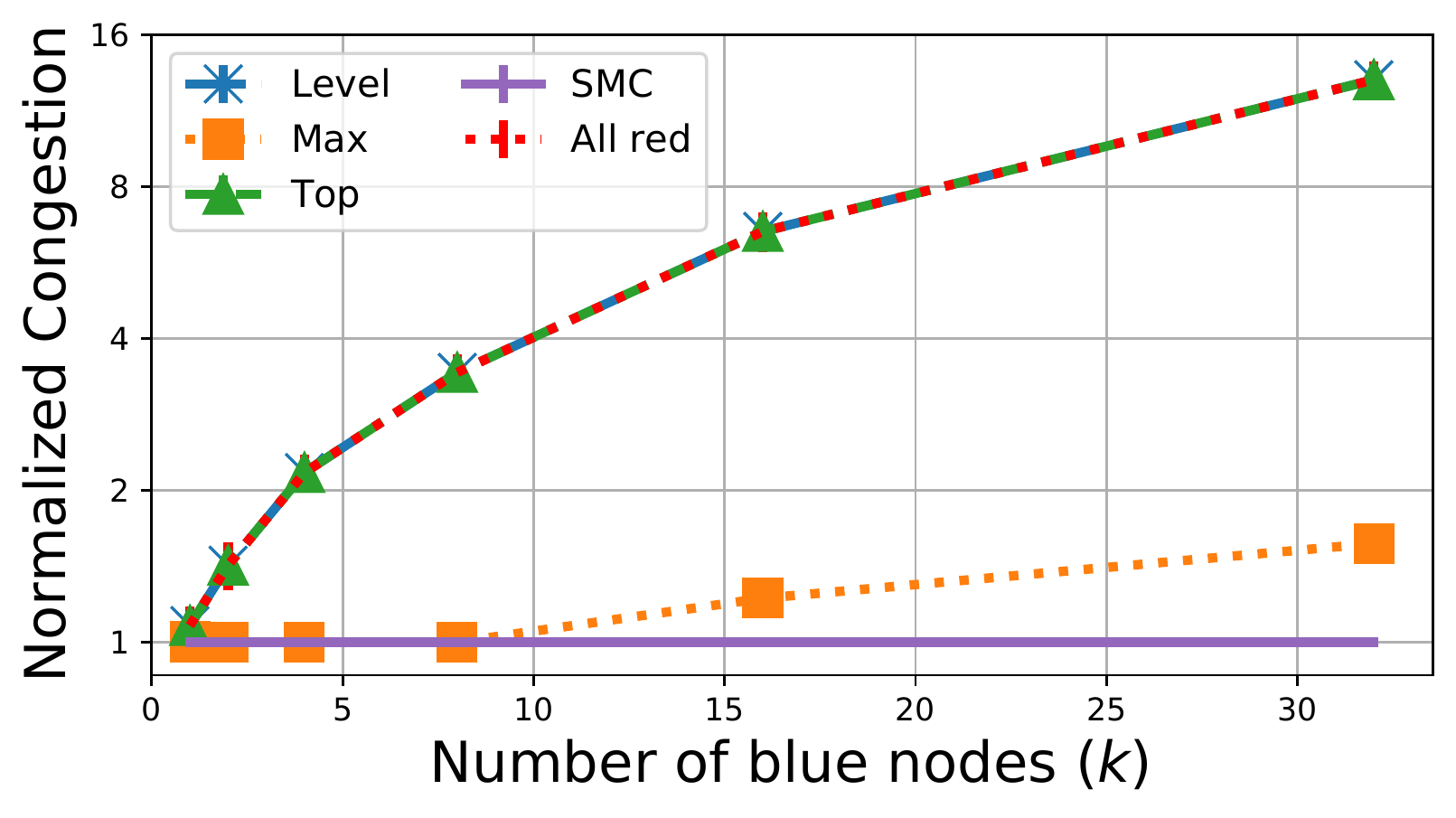} \\
        \raisebox{.35cm}{\rotatebox[origin=lc]{90}{\small Uniform load dist.}} \hspace{0.1cm} &
        \subcaptionbox{constant ($\weight=1$) \label{fig:VS_constant}}{
            \includegraphics[width=0.29\textwidth]{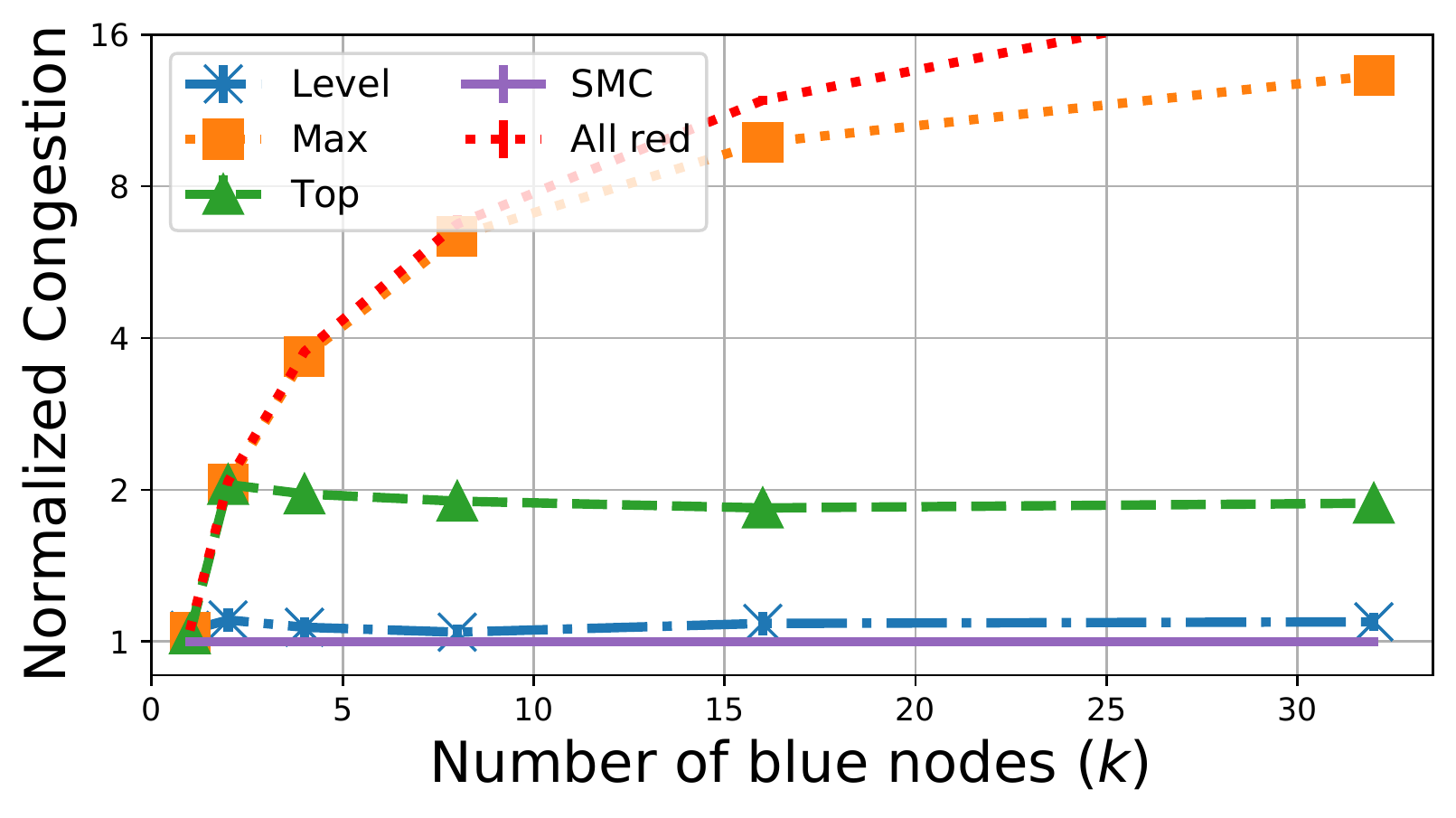}
        }&
        \subcaptionbox{linear increasing ($\weight=i$) \label{fig:VS_linear}}{
            \includegraphics[width=0.29\textwidth]{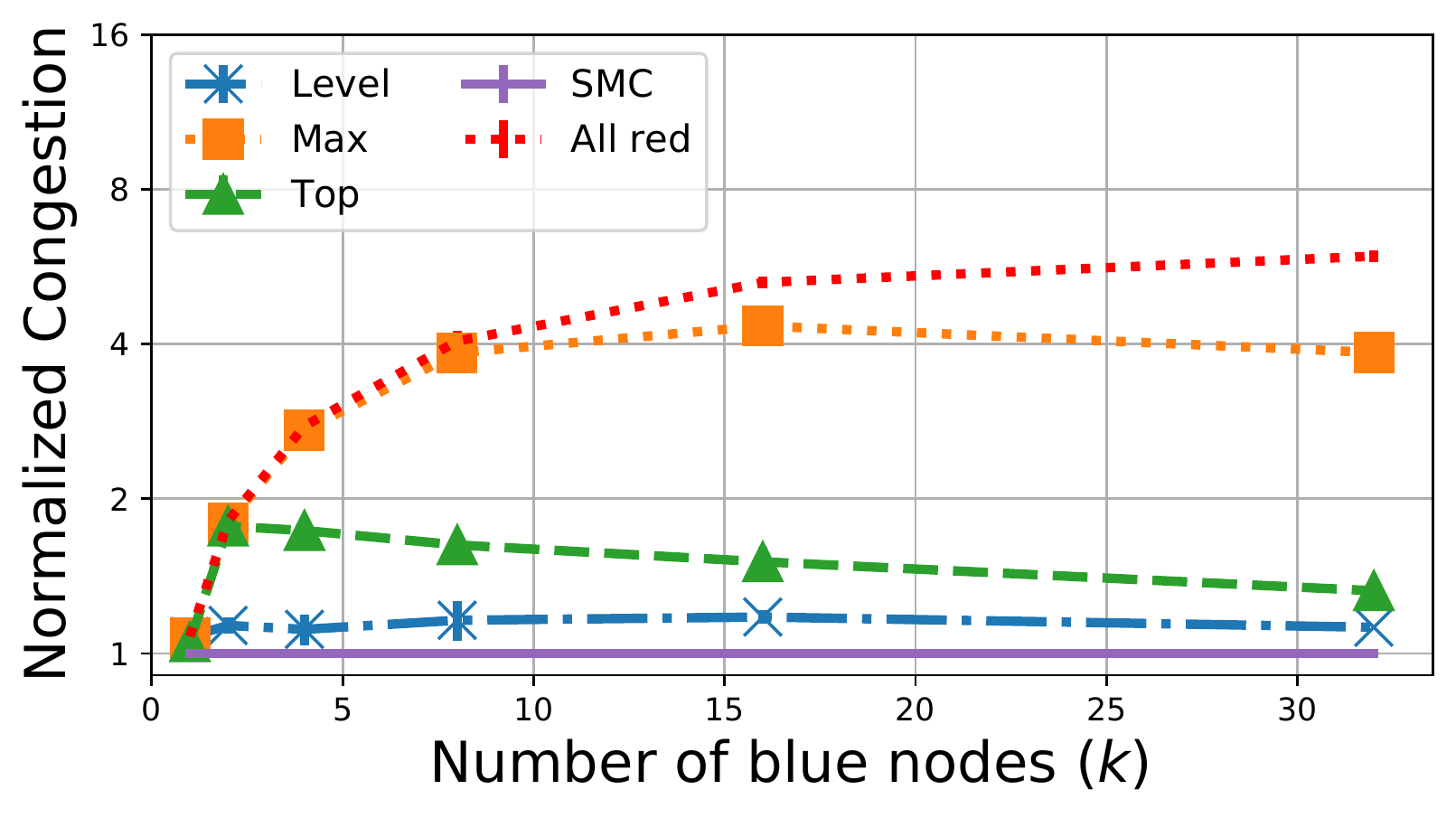}
        }&
        \subcaptionbox{exponentially increasing ($\weight=(1.5)^i$) \label{fig:VS_exponential}}{
            \includegraphics[width=0.29\textwidth]{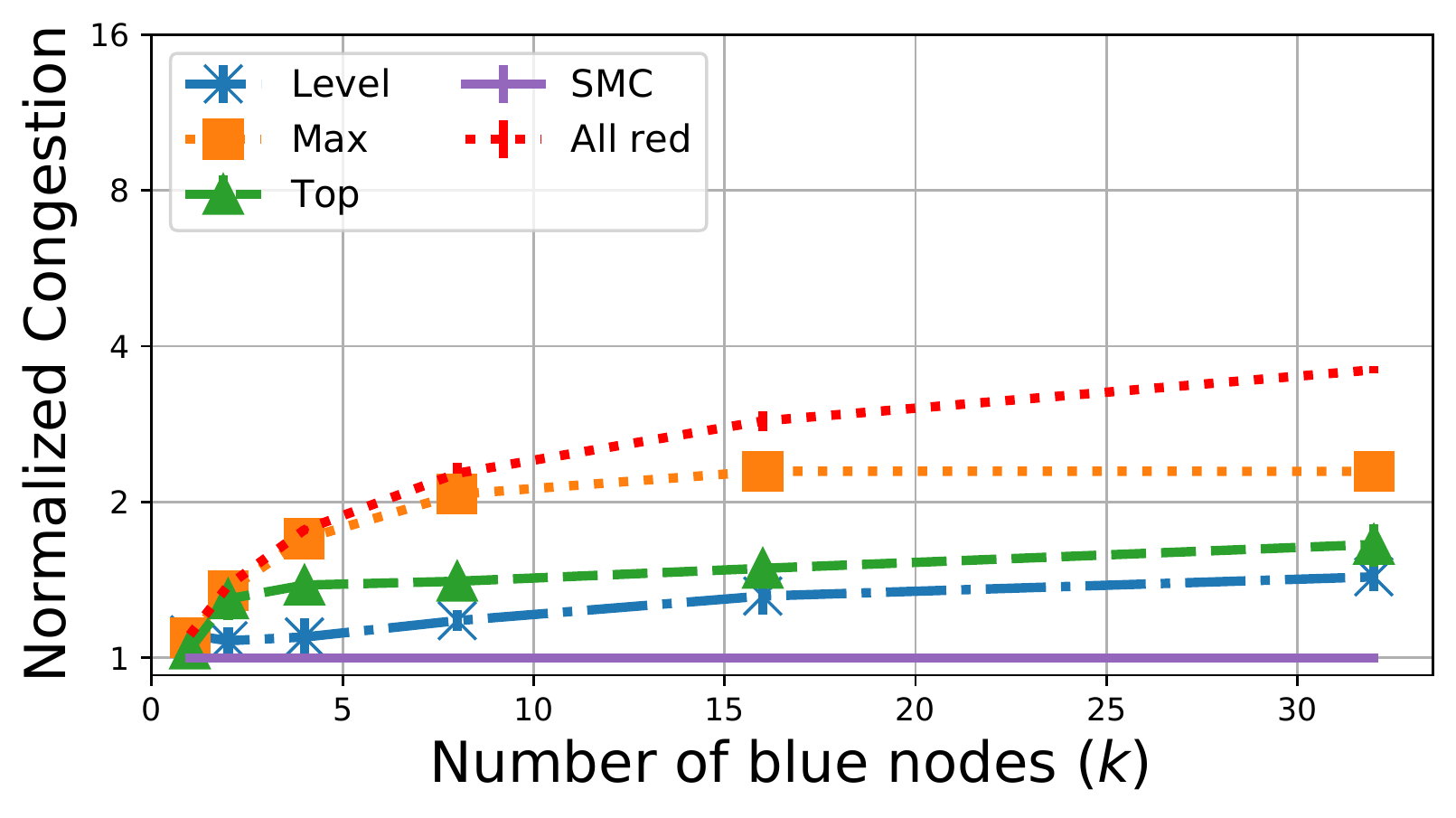}
        }
    \end{tabular}
\caption{\alg\ vs. other strategies for distinct schemes of rates (Fig.~\ref{fig:VS_constant}-\ref{fig:VS_exponential}), and distinct load distributions (power-law in the top plot, uniform in the bottom plot).}
\label{fig:VS}
\end{figure*}


\section{Evaluation}
\label{sec:evaluation}

In this section we report the results of our extensive evaluation of \alg. Our results shed light on various aspects pertaining to its performance, and also on the problem it is designed to solve.
In our evaluation, we examine both the network congestion induced by \alg, as well as that obtained by contending strategies.
We also show the result of running distributed application, including {\em word count} using the MapReduce paradigm, and gradient aggregation in distributed machine learning. These results essentially perform the Reduce operation on real workloads, thus highlighting real-world benefits.

We use the following setup for most of our evaluation (unless explicitly stated otherwise).
Our network is a complete binary {\em tree} with 255 nodes (and 128 leaves), where links have weights denoting their capacity.
We place load only in the {\em leaves} of the tree, which serve as {\em top-of-the-rack (ToR) switches} connected to servers (workers) that generate load.
The remaining network switches model the higher levels of a datacenter network, which facilitates a flow of information from the worker to the destination, serving as the aggregation server, that is connected to the root of the tree.

We consider two distributions for the load generated at the leaves, both with an average load of $5$ workers per ToR switch:
\begin{inparaenum}[(i)]
\item an almost {\em uniform} load, where the load of each node is picked u.a.r. in the range of integers $[1,9]$ (with variance $2.6$), and
\item a {\em power-law} load, where the (integer) load of each node is picked from a power-law distribution in the range $(1,63)$ (with variance $97.1$). 
\end{inparaenum}

We further consider three different rate schemes for the links in the tree:
\begin{inparaenum}[(i)]
\item {\em constant} rates, were all link rates are equal to $1$,
\item {\em linear} rates, were $\weight(e)$ increases linearly, by adding $1$, from leaf edges (rate $1$) towards the root, with a maximum rate of 7 in links entering the root, and
\item {\em exponential} rates, were $\weight(e)$ increases 
exponentially with base $1.5$, from leaf edges (rate $1$),  towards the root, with a maximum rate of 17 in links entering the root.
\end{inparaenum}

Each experiment was repeated ten times and we present the average performance for each such set of experiments. For clarity we present error bars only where we encountered significant variance in the results.

\subsubsection*{\bf The gains from limited In-network aggregation }

We first consider the network congestion reduction when using limited in-network aggregation resources.
Fig. \ref{fig:gains} presents the network congestion of \alg\ for the three rate schemes and the two distinct workload distributions, where the number $\numblue$ of blue nodes we are allowed to use takes values in $\numblue= 1, 2, 4, 8, 16, 32$. 
The figure also shows the network congestion for the {\em all-blue} and the {\em all-red} scenarios, which provide upper- and lower-bounds on the possible congestion.

The main takeaway from this figure is that in-network aggregation reduces the network congestion, and does that at a fast pace; Even with a small number of aggregation switches a significantly reduction is achieved.
Specifically, in all cases using merely 32 aggregation switches, which are about 12\% of the nodes, induces a x10 reduction in network congestion, which is close to the congestion obtained in the all-blue scenario.

\subsubsection*{\bf Comparing \alg\ with Other Strategies}
\label{sec:evaluation:compare_algorithms}

We now consider the performance of \alg\ compared to the performance of several contending strategies for solving the \combic\ problem.
Specifically, we focus our attention on the simple strategies described in our motivating example in Sec.~\ref{sec:example}, namely,
\begin{inparaenum}[(i)]
\item {\em \topalg},
\item {\em \maxalg}, and
\item {\em \levelalg}.
\end{inparaenum}

Fig. \ref{fig:VS} presents the performance of \alg\ alongside the performance of the contending strategies in the three rate scheme (left to right), 
for the two different workload distribution (top and bottom), where we consider $\numblue=1,2,4,8,16,32$.
and the network congestion of each algorithm  is {\em normalized} to the network congestion achieved by our algorithm, \alg, which was shown to be optimal in Sec.~\ref{sec:analysis}.
We further plot the performance of the {\em all-red} solution for reference.
As would be expected (by the optimality of \alg), all strategies preform worse then \alg, sometimes as much as x13 worse.

\begin{figure*}
    \centering
    \begin{tabular}{cccc}
        \subcaptionbox{constant ($\weight=1$) \label{fig:multiple_workloads_constant}}{
        \includegraphics[width=0.29\textwidth]{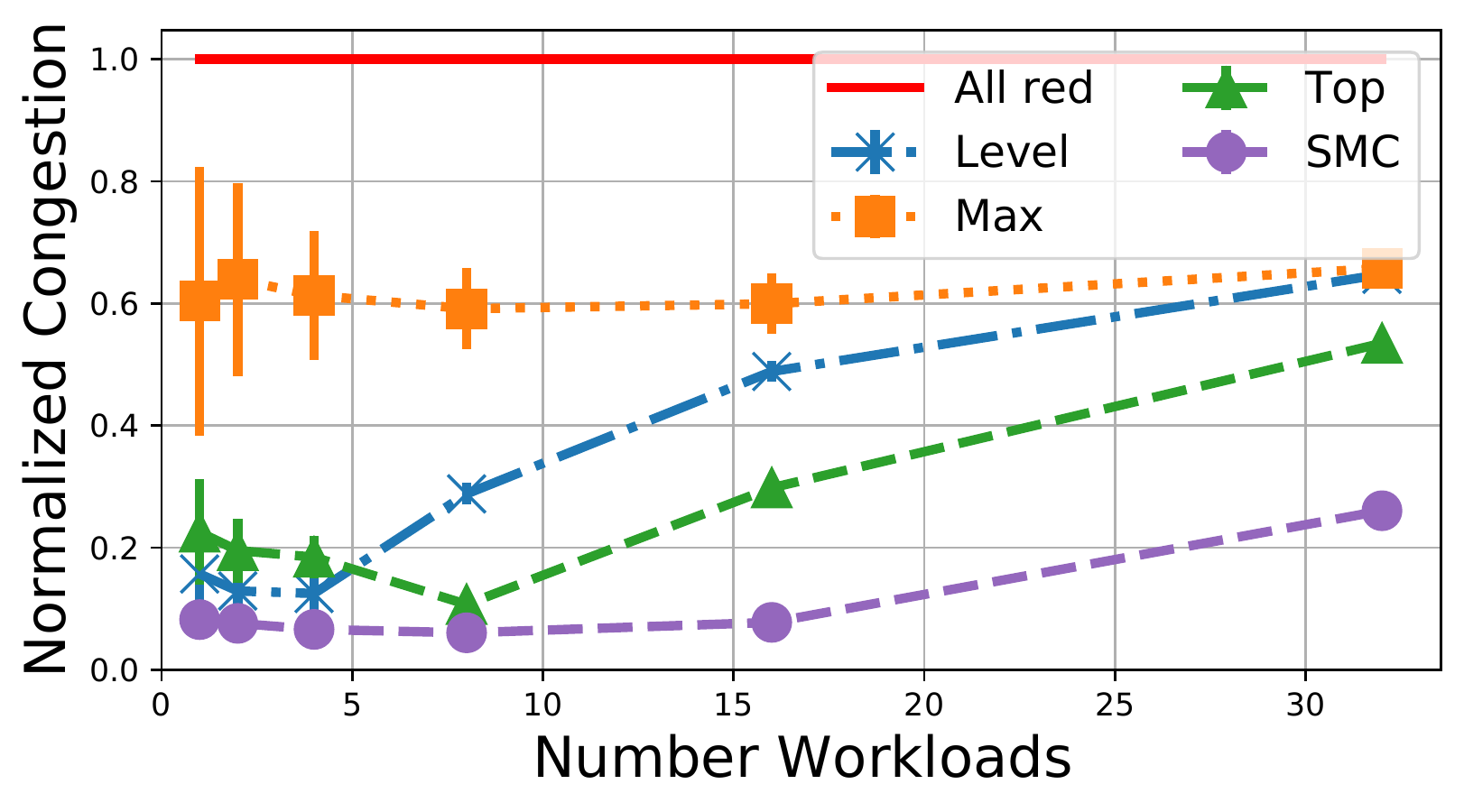}} &
        \subcaptionbox{linearly increasing ($\weight=i$) \label{fig:multiple_workloads_linear}}{
        \includegraphics[width=0.29\textwidth]{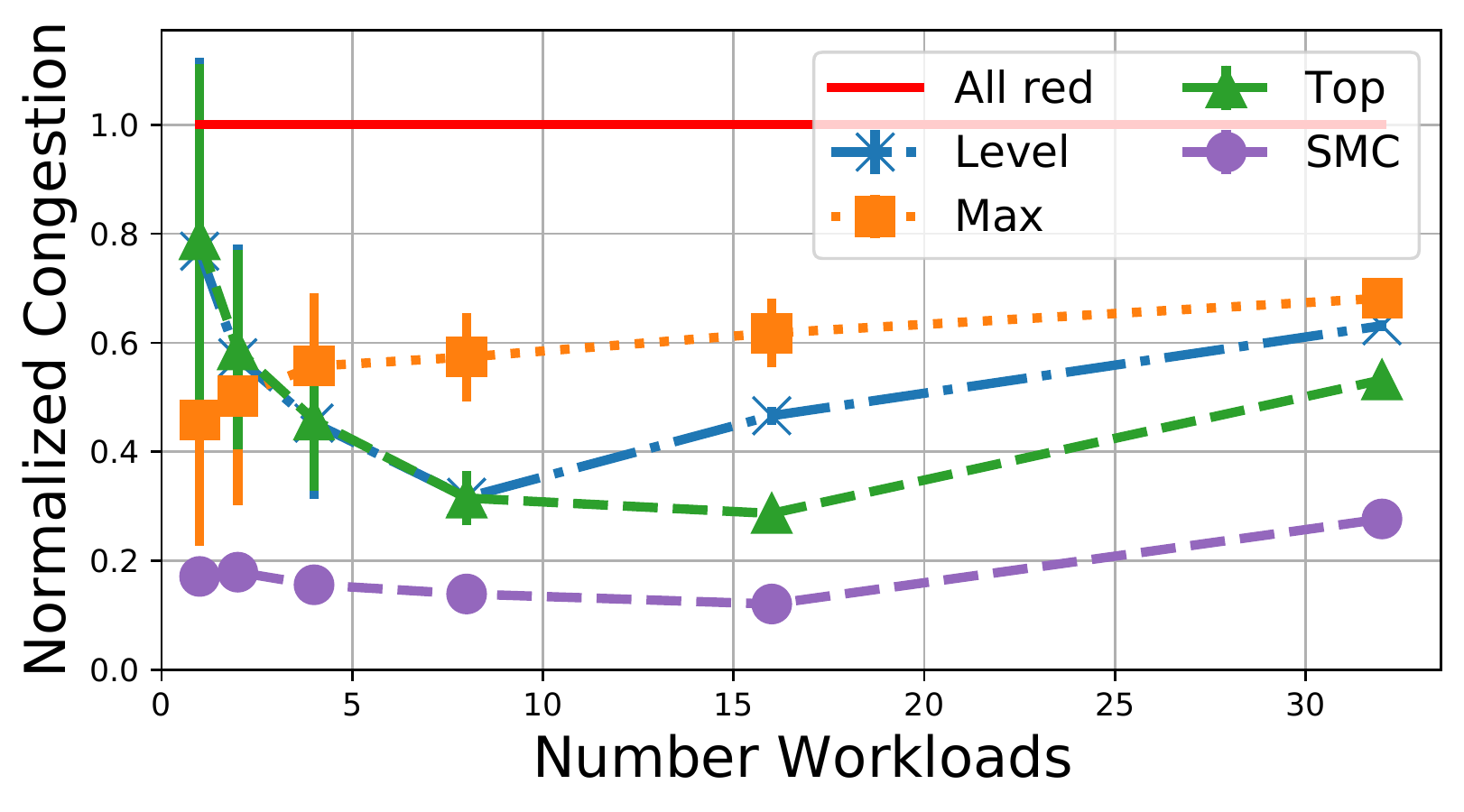}}&
        \subcaptionbox{exponentially increasing ($\weight=(1.5)^i$) \label{fig:multiple_workloads_exponential}}{
        \includegraphics[width=0.29\textwidth]{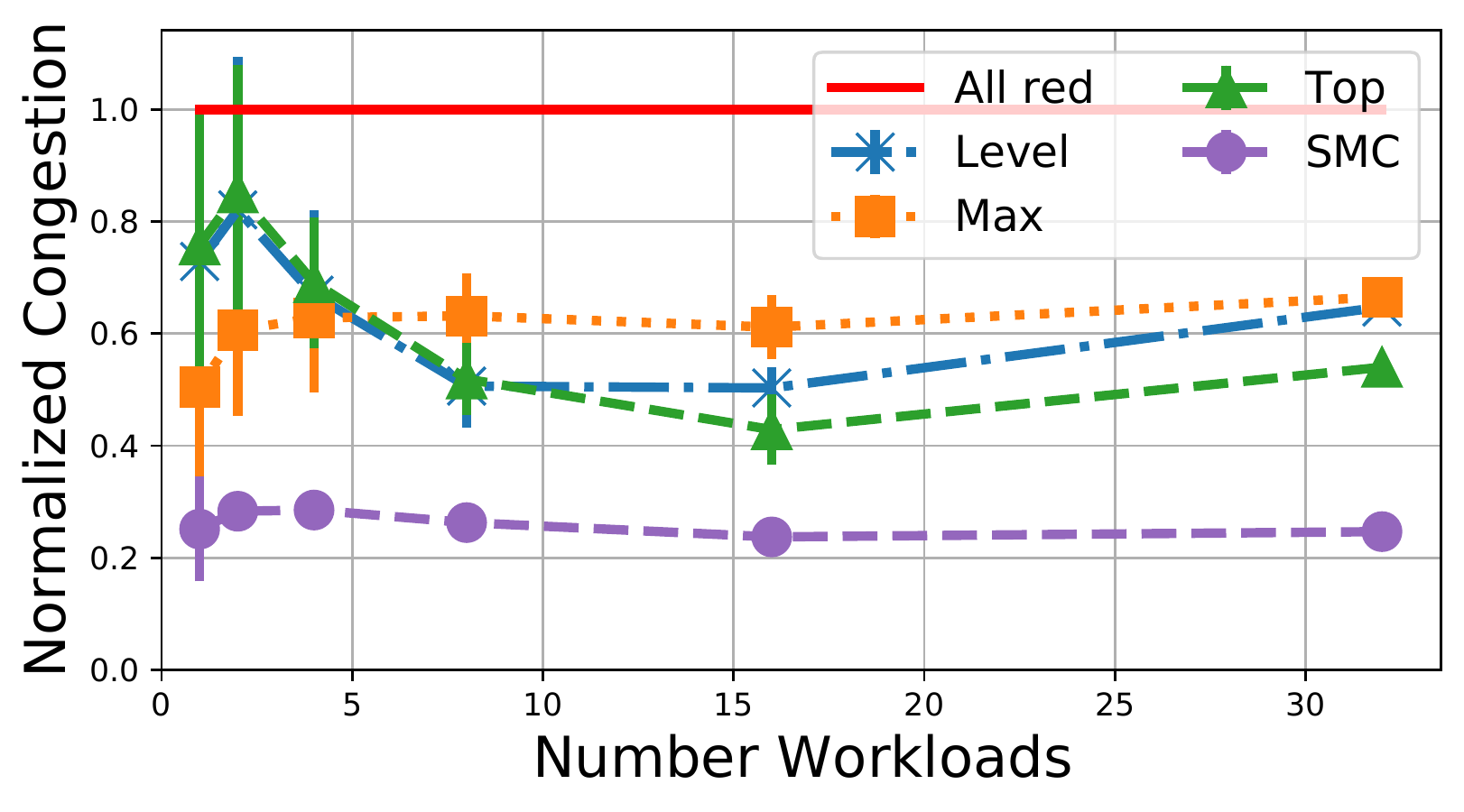}}
    \end{tabular}
\caption{\alg\ vs. other strategies when aggregating increasing the number of workloads. The switch aggregation capacity is fixed at $4$, and the $\numblue=16$}
\label{fig:multiple_workloads}
\end{figure*}

One can note that with the power-law workload distribution, and with constant rates, \maxalg\ performs worse than \topalg\ and \levelalg\ (\ref{fig:VS_constant}, top), while for the linear and exponentially increasing rates it outperforms them (\ref{fig:VS_linear} and \ref{fig:VS_exponential}, top).
This is due to the {\em location} where maximum link congestion is encountered.
In the constant rate regime the maximum link congestion occurs closer to the root of the tree. In contrast, when link rates are higher, the maximum congested link is ``pushed'' farther from the root, towards the leaves.
However, this phenomena does not assist \maxalg\ under the {\em uniform load distribution}, since, due to the smaller variance of this distribution, \maxalg\ is unable to reduce all heavily loaded ToR switches.

Since \alg\ is optimal, it exhibits the best performance in all scenarios.
This serves to show that using \alg\ ensures robustness regardless of load distribution or link rates.
However, the second-best strategy strongly depends on the load distribution, or the link rates.
The power-law load distribution favors the \maxalg\ strategy, since high-load ToR switches that perform aggregation induce a significant reduction in congestion.
For the uniform distribution, however, the \levelalg\ strategy fares best, since it manages to load balance the uniform loads at the leaf-switches throughout the network. 
The \topalg\ strategy is the most sensitive to the link rates, where having higher rates towards the root of the network implies that performing in-network aggregation further up
provides very little benefits compared to performing aggregation closer to the leaves.

\subsubsection*{\bf Multiple Workloads}
\label{sec:evaluation:multiWL}

We now turn to address the problem of handling {\em multiple workloads}, and determining where aggregation should take place for each such workload. We note that this serves as an extension of our framework that goes beyond the model described in Sec.~\ref{sec:model}.
Each workload $\load_t$ is determined by its time, $t=0,1,2,\ldots$.
We consider a sequence of workloads, $\load_t$, $t=0,1,2,\ldots$, arriving in an {\em online} fashion, such that determining the aggregating switches for workload $\load_t$ should be settled before handling workload $\load_{t+1}$.

We further assume each switch $\switch$ has a predetermined {\em aggregation capacity} $\aggcap(\switch)$ which bounds the number of workloads for which $\switch$ can be assigned as an aggregating switch.
We let $\aggcap_t(\switch)$ denote the {\em residual aggregation capacity} remaining at $\switch$ before handling workload $\load_t$.
If switch $\switch$ is designated as an aggregation switch when handling workload $\load_t$, then $\aggcap_{t+1}(\switch)=\aggcap_t(\switch)-1$, and $\aggcap_{t+1}(\switch)=\aggcap_t(\switch)$ otherwise.

We examine the performance of the various strategies considered in Sec.~\ref{sec:evaluation:compare_algorithms},
when applied repeatedly to the sequence of workloads $\load_0, \load_1, \ldots$, given as input.
The set of switches available for aggregation when handling workload $\load_t$ is defined by $\Availability_t=\set{\switch \mid \aggcap_t(\switch) > 0}$.

We generate our sequence of workloads in an online fashion, by drawing each workload from either the uniform load distribution, or the power-law load distribution, each with probability $\frac{1}{2}$, and use as our baseline the values
$k=16$ and $\aggcap(\switch)=4$ for every switch $\switch$. We evaluate the system's performance when handling more and more workloads, where we specifically consider handling $1, 2, 4, 8, 16, 32$ workloads.

Fig.~\ref{fig:multiple_workloads} shows the performance of \alg\ compared to the performance of the various strategies described in Sec.~\ref{sec:example}.
Similarly to our previous results,
our evaluation considers 3 scaling laws for link rates: constant (in Fig.~\ref{fig:multiple_workloads_constant}), linearly increasing (in Fig.~\ref{fig:multiple_workloads_linear}), and exponentially increasing (in Fig.~\ref{fig:multiple_workloads_exponential}).

The figure shows the normalized network congestion, where normalized to the congestion obtained by the all-red solution.
Namely, if the performance of an algorithm is $\alpha \in [0,1]$ in some scenario, this means that the algorithm entails a network congestion that is an $\alpha$ fraction of the congestion incurred by the all-red scheme.
Notice that as the number of workloads increases, the performance of any strategy would converge to that of the all-red configuration.
This follows from the fact that the aggregation capacity is bounded, implying that once the number of workloads is large enough, further workloads cannot benefit from any aggregation, and the initial benefits of aggregating the prefix of the workload arrival sequence become marginal compared to the toll imposed by the entire sequence.
This explains the worsening performance exhibited 
when increasing the number of workloads.
Nevertheless, for the exponential rates regime \alg\ is able to sustain a larger amount of workloads before changing for the worse.

\subsubsection*{\bf Switch Capacity}
\begin{figure}
    \centering
            \includegraphics[width=0.31\textwidth]{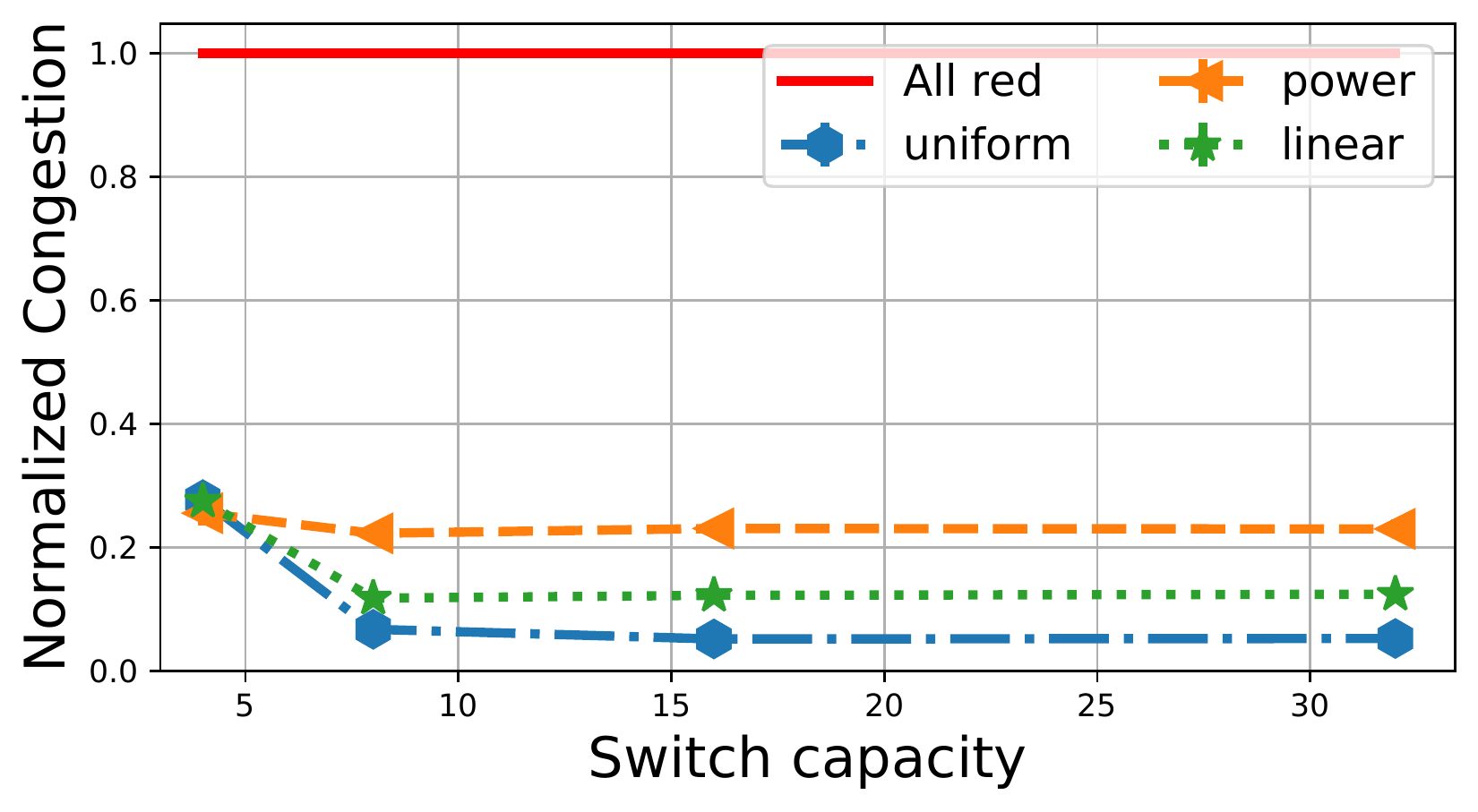} 
\caption{\alg\ performance when increasing the switch capacity, $32$ workloads and $\numblue=16$ per workload.}
\label{fig:switch-capacity}
\end{figure}

We now turn to evaluate the effect of the switch in-network capacity.
Similarly to section~\ref{sec:evaluation:multiWL} we normalized the results to the all-red scenario, and consider distinct link rates environments.

Fig~\ref{fig:switch-capacity} shows the effect of varying the aggregation capacity on the performance of \alg, while using $k=16$, 32 workloads, and distinct values $\aggcap(\switch)=4,8, 16, 32$ for every switch $\switch$.
In such a scenario, clearly a capacity of 32 will yield the best performance, as capacity is abundant, and each workload can be aggregated optimally, independently of other workloads.
However, as shown in fig~\ref{fig:switch-capacity}, \alg\ actually achieves this optimal performance with significantly smaller switch capacity.

\subsubsection*{\bf \alg\ for Different Applications}
\label{sec:evaluation:compare_applications}


We now consider two use cases for evaluating the system:
\begin{inparaenum}[(i)]
\item {\em big-data}, using a word-count task~\cite{apache}, where we make use of a wikipedia dump~\cite{wiki}, with an overall of 54M words, out of which 800K are unique. We refer to this use case as the {\em word count (\wcapp)} use case.
\item {\em distributed ML}, using distributed gradient aggregation with a parameter server~\cite{li14scaling}, where worker servers independently perform neural-network training, over a 10K feature space, using 0.5 dropout rate~\cite{srivastava2014dropout}.
The workers send their updated gradients to a parameter server, which then updates the system model parameters.\footnote{We note that our work considers solely the network congestion produced by such tasks, and not the quality of the model produced, which may depend on a variety of problem characteristics. We therefore do not implement the actual neural network, but rather consider the messages sent by the worker servers, and the aggregation of these messages.}
We refer to this use case as the {\em parameter server (\psapp)} use case.
\end{inparaenum}

We evaluate the performance of \alg\ for \wcapp, and \psapp, using the constant rates regime, which better highlights the differences in the performance, and using the uniform distribution which is more challenging for reducing congestion.

\begin{figure}
    \centering
            \includegraphics[width=0.35\textwidth]{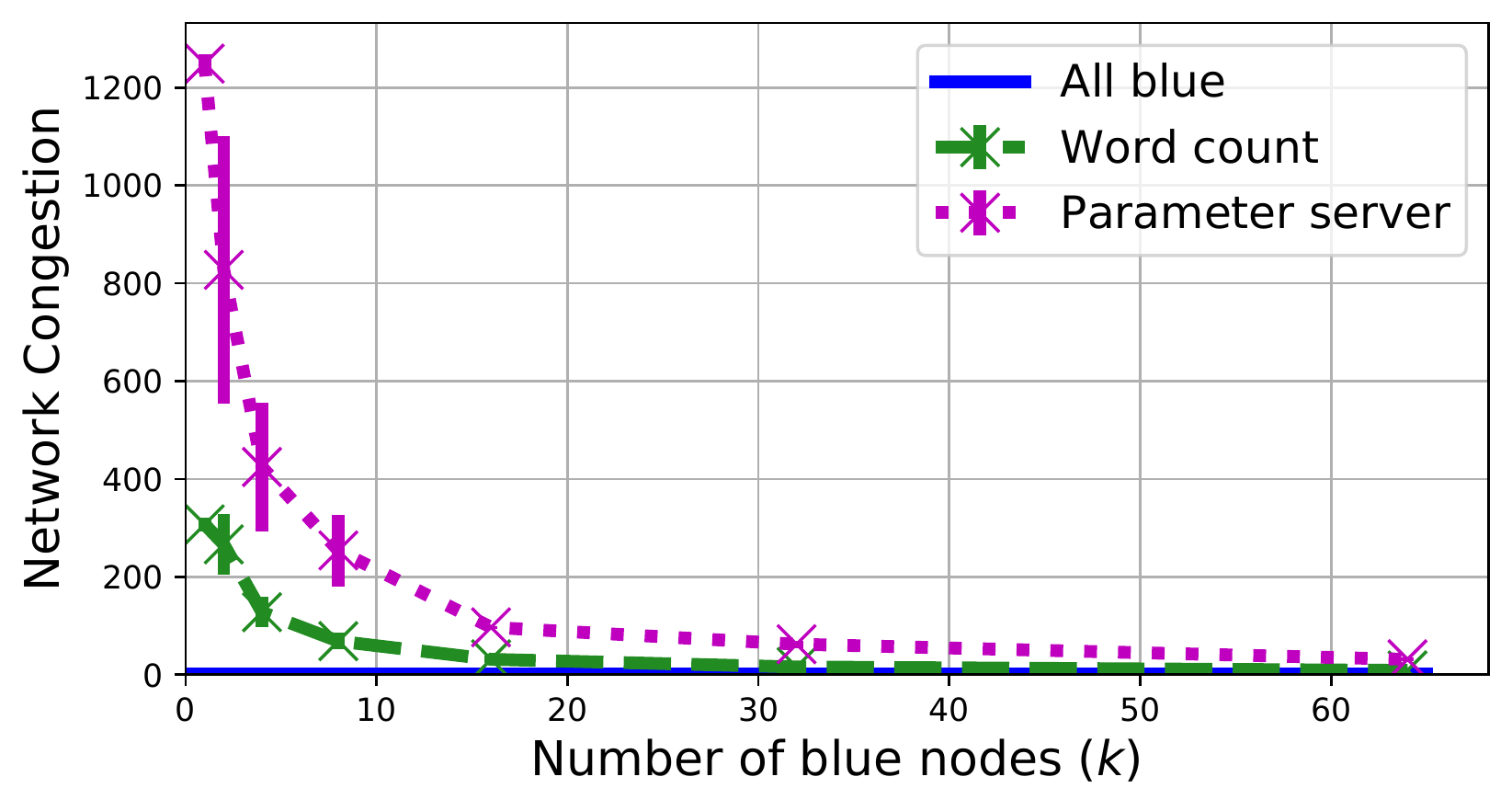}
\caption{\alg\ performance for the \wcapp\ and \psapp\ use cases.}
\label{fig:apps}
\end{figure}

Fig.~\ref{fig:apps}
shows
the results of our evaluation, where the congestion attained by \alg\ is normalized to that of the all-red scenario.
This figure highlights the significant reduction in network congestion even when using a small number of aggregation switches.
The main takeaway here is that the {\em application} scenario has a significant impacts on the perceived network congestion.
While in the \psapp\ use-case the congestion is very high without aggregation and rapidly improves once (limited) aggregation is deployed, for the \wcapp\ use-case network congestion is significantly smaller apriori, and the improvement obtained by deploying few aggregation switches is milder.

\section{Related Work}
\label{sec:relatedwork}
Various studies considered data aggregation~\cite{jesus15survey}, covering diverse domains such as wireless networks, scheduling, etc.~\cite{nakamura07information,malhotra11aggregation}, and studying which functions may be aggregated efficiently~\cite{yu09distributed,jesus15survey}.
Furthermore, as discussed in Sec.~\ref{sec:introduction}, data aggregation is a cornerstone of big data tasks, using, e.g., the MapReduce framework~\cite{dean04mapreduce,mai14netagg}, and more recently also of distributed machine learning (ML) environments, performing, e.g., the training of deep neural networks.

Specifically for such ML tasks, network performance has been noted as a major bottleneck hindering the efficient usage of such frameworks~\cite{li14communication,viswanathan20network}.
Various approaches have been suggested to modify ML methodologies in order to improve upon the network induced performance of distributed ML~\cite{xu20compressed,dutta20discrepancy,wang20geryon}. Additional network- and system-level adaptations have been suggested to improve upon ML performance of such systems~\cite{abdelmoniem21impact,wang19impact,ouyang21communication}.
A notable use-case which applies to our framework is the usage of a {\em parameter server} for aggregating and distributing model parameters~\cite{li14scaling}, where various works addressed the networking overheads it entails~\cite{li14communication,mai15optimizing,luo18parameter}.
Additional approaches focus on {\em gradient aggregation}, where merely gradients are aggregated and distributed to the workers. This concept has gained significant popularity in frameworks of federated ML~\cite{reisizadeh19robust}.
A special emphasis is notably given for supporting large scale ML in High-Performance Computing (HPC) clusters, including specially tailored protocols for doing in-network aggregation (e.g., nvidia's SHARP~\cite{graham20sharp}).


More generally, in-network computing has been the focus of much attention, fueling the design of advanced architectures ranging from network HW design~\cite{eran19nica}, through networking services~\cite{shantharama20hardware}, up to various applications~\cite{dang20p4xos,tokusashi18lake,vaucher20zipline}, including ML\cite{sapio19scaling,gebara21innetwork}, to name but a few.

We note that the majority of these work address the incorporation of specific functionalities within the network, or the application.
In contrast, our work considers a more general network-level problem focusing on resource allocation and placement within the network, in scenarios where resources are scarce, in an attempt to optimize system performance, independent of the specific application being served.

\section{Discussion and Future Work} 
\label{sec:disussion_future_work}

This work considers the \combic\ problem, where we need to determine the location of a limited number of aggregation switches performing a reduce operation, within a tree network, so as to minimize the network congestion.
This problem lays at the heart of many distributed computing use cases, and most notably in variations of the {\em AllReduce} operation for distributed and federated machine learning.
Our work describes an optimal algorithm, \alg, for solving the \combic\ problem in trees, and provides insights as to the performance of \alg\ via an extensive simulation study.


Developing solutions that are applicable to {\em general} networks (i.e., not necessarily tree networks), thus supporting multi-path routing is a challenging task we leave for future research.
Obtaining worst-case guarantees for multiple workloads is another interesting open problem.
The main challenge there is how to distribute remaining aggregation capacity throughout the network to the various workloads.
In general, we may serve every workload using a {\em different} number of aggregation switches (i.e., there need not be a uniform $k$ for all workloads).
Finally we would like to target minimizing the {\em delay} incurred by the system,
and we expect our general algorithmic approach to also be effective for such objectives. 

\label{end_body}

\clearpage

\bibliographystyle{IEEEtran}
\bibliography{bibliography}
\end{document}